\documentclass{JHEP3}
\usepackage{amssymb,amsmath}
\usepackage{epsfig}
\usepackage{citesort}

\title{Dark energy as a mirage}

\author{Teppo Mattsson$^{1,2,}$\footnote{E-mail: forename.surname@helsinki.fi}\\
${}^1$ Helsinki Institute of Physics, P.O. Box 64, FIN-00014 University of Helsinki, Finland\\
${}^2$ Department of Physical Sciences, P.O. Box 64, FIN-00014
University of Helsinki, Finland}

\abstract{Motivated by the observed cosmic matter distribution, we present the following conjecture: due to the formation of voids and opaque structures, the average matter density on the path of the light from the well-observed objects changes from $\Omega_M \simeq 1$ in the homogeneous early universe to $\Omega_M \simeq 0$ in the clumpy late universe, so that the average expansion rate increases along our line of sight from EdS expansion $Ht \simeq 2/3$ at high redshifts to free expansion $Ht \simeq 1$ at low redshifts. To calculate the modified observable distance-redshift relations, we introduce a generalized Dyer-Roeder method that allows for two crucial physical properties of the universe: inhomogeneities in the expansion rate and the growth of the nonlinear structures. By treating the transition redshift to the void-dominated era as a free parameter, we find a phenomenological fit to the observations from the CMB anisotropy, the position of the baryon oscillation peak, the magnitude-redshift relations of type Ia supernovae, the local Hubble flow and the nucleosynthesis, resulting in a concordant model of the universe with 90\% dark matter, 10\% baryons, no dark energy, $15~{\rm Gyr}$ as the age of the universe and a natural value for the transition redshift $z_0=0.35$. Unlike a large local void, the model respects the cosmological principle, further offering an explanation for the late onset of the perceived acceleration as a consequence of the forming nonlinear structures. Additional tests, such as quantitative predictions for angular deviations due to an anisotropic void distribution and a theoretical derivation of the model, can vindicate or falsify the interpretation that light propagation in voids is responsible for the perceived acceleration.}

\preprint{HIP-2007-64/TH}


\begin{document}

\section{Introduction}\label{intro}

The widely adopted framework of modern cosmology is the spatially homogeneous and isotropic Friedman-Robertson-Walker spacetime, with the growth of structure described as linear perturbations evolving on the smooth FRW background. Accordingly, the standard model of cosmology is built on the assumption that the effects of the evident nonlinear inhomogeneities on the detectable light average out\footnote{Apart from gravitational lenses that are occasionally taken into account as additional corrections.} over cosmological distances. This assumption -- although perhaps in concordance with Newtonian intuition -- lacks a convincing demonstration within general relativity, not to mention a mathematically rigorous proof. In fact, the assumption was questioned already in the 60's by Zel'dovich \cite{Zeldovich}, Feynman\footnote{Apparently, R. P. Feynman was also one of the pioneers of this subject, as \cite{Gunn1,Kantowski1} cite a colloquium by Feynman in 1964.}, Bertotti \cite{Bertotti}, Gunn \cite{Gunn1}, Kantowski \cite{Kantowski1} and in the 70's by Dyer \& Roeder \cite{DyerRoeder1,DyerRoeder2}, who suggested that the clumpiness of matter in the real universe has consequences on the observable distance-redshift relations that the simplified FRW description fails to capture.

At the time of this pioneering research, the cosmological observations were too inaccurate to distinguish the predictions of the proposed inhomogeneous models from the simple and elegant FRW solutions. Consequently, the efforts towards a more thorough description of the effects of the nonlinear structures on the distance-redshift relations never became part of the mainstream research. Although the structure formation itself has since then developed into a major part of modern cosmology, the emphasis is usually given either to the general relativistic but perturbative description of structures on large scales, or to the Newtonian description of the small scale structure e.g.\ in the context of galaxy formation.

At present, the unparalleled precision of the recent cosmological observations has raised the debate about the effect of the nonlinear structures on the distance-redshift relations once again topical. Indeed, the observations indicate the existence of inhomogeneities that seem to be on large enough scales to have cosmological significance but too strong to be described within the linear theory --- in particular the observed voids and the clustering of matter as filaments of opaque galaxies between the voids \cite{Hoyle:2003hc,Gott:2003pf,Tikhonov:2007di,vonBendaBeckmann:2007wt,Einasto,Rudnick:2007kw}. The increase in the precision of the data naturally demands a corresponding improvement in the theoretical model; otherwise there is an increased risk of drawing incorrect conclusions from the observations.

In fact, the discovery that the standard FRW cosmology needs a fine-tuned dark energy component \cite{Copeland:2006wr,Straumann:2006tv,Sahni:2006pa} in order to account for observations \cite{Riess:2004nr,Riess:2006fw,Eisenstein:2005su,Spergel:2006hy}, suggests that the precision of the data may already have surpassed the precision of the employed theoretical model. This is also backed up by a mysterious coincidence, pointed out by Schwarz \cite{Schwarz:2002ba} and Wetterich \cite{Wetterich:2001kr} and elaborated by R\"as\"anen \cite{Rasanen:2003fy,Rasanen:2006kp,Rasanen:2008it,Rasanen:2008be}: the perceived increase in the cosmic expansion, ascribed to dark energy in the standard cosmology, happens to take place during the same era when nonlinear structures start to form at cosmologically significant scales.

Whereas the fine-tuning of dark energy only indicates that something fundamental may be lacking from the standard picture, the coincidence problem suggests a causal connection between cosmic acceleration and structure formation. Perhaps the most natural explanation would be, that the standard FRW description breaks down when the nonlinearities become dominant in the late universe and the need for a tiny cosmological constant is a manifestation of this breakdown, not evidence for dark energy \cite{Rasanen:2003fy}. In recent years, more and more authors have addressed this issue and substantially increased the general understanding of the subject, but crucial open questions still remain \cite{Buchert:2007ik}. In this work, we address the issue by considering a phenomenological model that aims to: 1) clarify the physical interpretation for \emph{how} the structure formation can actually mimic dark energy, and 2) provide a quantitative fit for the observations.

The paper is organized as follows. In Sects.\ \ref{noijannuoli}--\ref{selectioneffects}, we review three physically independent mechanisms for how the cosmic inhomogeneities can induce accelerated expansion along our line of sight: global acceleration due to spatial variations in the expansion rate, faster local expansion rate due to a large local void, and biased light propagation through voids that expand faster than the average; possibilities to describe the total effect of the inhomogeneities have been discussed in Sect.\ \ref{TotalEffect}. After reviewing the original Dyer-Roeder method in Sect.\ \ref{originalDR}, and motivating the need to modify it, we introduce a generalized Dyer-Roeder method in Sect.\ \ref{dyerroeder} to allow for the total effect of the inhomogeneities on the observable distance-redshift relations. The distance-redshift relations are then calculated for an inhomogeneous model, based on the conjecture that the average matter density on the path of the light from the observed objects changes from $\Omega_M \simeq 1$ in the homogeneous early universe to $\Omega_M \simeq 0$ in the clumpy late universe. The physical foundations of the conjecture are summarized in Sect.\ \ref{physics}. In Sect.\ \ref{derivation}, we construct a simple description of the structure formation in terms of a function $\alpha(z)$ that measures the deviation of the matter density along our line of sight from the FRW matter density. A comparison of the model with cosmological observations is performed in Sect.\ \ref{observations}, and a discussion of the results is given in Sect.\ \ref{results}. Finally, Sect.\ \ref{konkluusiot} contains our conclusions.

\section{Inhomogeneous generalizations of the FRW cosmology}\label{classification}

Formally, the cosmic inhomogeneities pose the following problem: we have a physical system, the universe, with $>10^{80}$ degrees of freedom but we want a model with $\lesssim 10$ degrees of freedom, represented by the cosmological parameters. For a mathematical description of the universe, we thus need a coarse graining map: $10^{80}~{\rm{dof}} \mapsto 10~{\rm{dof}}$, that preserves the relevant physical structure of the spacetime but removes its mathematical complexity. As our cosmological knowledge is firmly based on observing light, the relevant structure is the correct optical properties of the universe. That is, we want the observable distance-redshift relations to be unaltered by the coarse graining.

A common critique against acceleration from inhomogeneities is that linearly perturbed FRW solutions seem to describe well even highly nonlinear density contrasts, $\delta \rho/ \bar{\rho} \sim 10^{30}$, while still having insignificant backreaction \cite{Ishibashi:2005sj}. This is because the relevant perturbed gravitational potential can satisfy $\Phi \ll 1$ even for the high density contrasts $\delta \rho/ \bar{\rho} \gg 1$. However, this argumentation overlooks the crucial part of realistic nonlinear structures, namely inhomogeneities in the expansion rate \cite{Rasanen:2006kp}. Even a relatively small spatial variation in the expansion rate, $\delta \theta /\bar{\theta} \sim {\mathcal{O}}(0.1)$, can invalidate the linear perturbation theory by rendering the uniformly expanding FRW background inapplicable. Sometimes the assumption $\delta \theta /\bar{\theta} \ll 1$ is hidden in the requirement of small peculiar velocities \cite{Paranjape:2008jc}, which appears to restrict the spatial variations in the expansion rate to be small as well \cite{Kolb:2008bn}. As we next discuss, ${\mathcal{O}}(0.1)$ spatial variations in the expansion rate can indeed induce accelerated expansion via various mechanisms.

\subsection{Backreaction due to inhomogeneous expansion}\label{noijannuoli}

A central question in constructing the coarse grained description of the universe is how to average relativistic gravitational systems within Einstein's theory. As pointed out by Shirokov and Fisher \cite{Shirokov} and later made more popular in the observational cosmology programme by Ellis and collaborators \cite{Ell84,EllisStoeger,Ellis:2005uz,Ellis:1999sx}, it seems physically more correct to first calculate the Einstein field $G(g)$ for the exact metric $g$ and only then average $\langle G(g) \rangle$, than to calculate the Einstein field for an averaged metric $G (\langle g \rangle)$ as in the standard FRW cosmology. The reason is that the Einstein field is more closely related to physical quantities whereas the metric corresponds to a gravitational tensor potential, whose derivatives determine the physics. In general relativity, the field $G$ depends nonlinearly on the metric $g$, so its evaluation does not commute with averaging and the non-vanishing difference, $\langle G(g) \rangle - G (\langle g \rangle ) \neq 0$, gives rise to what is known as the nonlinear \emph{backreaction}; see \cite{Rasanen:2003fy,Rasanen:2006kp,Rasanen:2008it}. Hence the issue is not only a conceptual one, but in general, the two approaches yield identical predictions only in the absence of nonlinear inhomogeneities.

Even if performed in the above explained order, the averaging is not free of problems. Firstly, the only practicable way to average tensors seems to be the manifestly coordinate-dependent averaging of components: $\langle G \rangle \equiv \langle G_{\mu \nu} \rangle$. As we would like the coordinate-invariance of general relativity to hold also for the coarse-grained or averaged quantities, only the inherently invariant rank $0$ tensors or scalars appear to have well-defined averages. Secondly, the desire in cosmology to divide the universe into temporal and spatial parts leads to the issue of foliation dependence: in order to take spatial averages, we must artificially break the symmetry between time and space inherent in general relativity.

In 1999, Buchert proposed a way to circumvent, or at least alleviate, these problems \cite{Buchert:1999er}. By using the fact that a part of the Einstein tensor equation can be decomposed into scalar equations, he derived a set of spatially averaged equations, now known as the Buchert equations, describing the averaged dynamics of a general irrotational dust universe. Moreover, he foliated the spacetime by using the proper time of the dust as a global time coordinate; a physically well justified choice in a dust universe, also supported by a comparison with exact observables performed in Ref. \cite{Mattsson:2007qp}.

To see how the backreaction can affect the observations, consider the Buchert acceleration equation \cite{Buchert:1999er}:
\begin{equation}\label{Buchert1}
3\frac{\ddot{a}}{a} = -4 \pi G \langle\rho\rangle +\mathcal{Q} ~,
\end{equation}
where $a(t)$ is a generalized scale factor, $\rho$ is the matter density and the difference between Eq.\ (\ref{Buchert1}) and its homogeneous FRW counterpart is known as the dynamical backreaction
\begin{equation}\label{backreaction}
\mathcal{Q} \equiv\frac{2}{3}(\langle\theta^2\rangle -\langle\theta\rangle^2 )- 2 \langle\ \hspace{-3pt} \sigma^2 \rangle ~,
\end{equation}
where $\sigma^2$ is the shear scalar and $\theta$ is the expansion scalar. The backreaction (\ref{backreaction}) in the Buchert equations explicitly demonstrate that averages of inhomogeneous quantities do not evolve in time like the corresponding homogeneous quantities.

From Eq.\ (\ref{Buchert1}) it is obvious that, regardless of decelerating locally everywhere, with large enough backreaction $\mathcal{Q} >4 \pi G \langle\rho\rangle $, the average expansion can accelerate without an exotic fluid with negative pressure or a modification of gravity. Physically, the global acceleration is possible because the volume can become dominated by fast-expanding regions \cite{Rasanen:2008it}. This is realized with large enough variance of the expansion rate, if the counterbalancing average shear is not too large. The variance becomes large when contracting ($\theta<0$) and expanding ($\theta>0$) regions coexist. This is exactly what one would expect in the late universe with structures forming via gravitational collapse \cite{Apostolopoulos:2006eg,Kai:2006ws}, so it has been conjectured that the average acceleration could explain the cosmological observations \cite{Rasanen:2006kp}.

To clarify the essential point, consider two disconnected decelerating regions $U$ and $O$ that have initially the same volumes $V_i$. Let the region $U$ represent an underdensity that initially expands at the rate $\theta_i>0$ and let $O$ represent an overdensity that initially contracts at the rate $-\theta_i<0$. Clearly, the volume-averaged expansion rate is initially zero: $\langle \theta \rangle_i = (V_i \theta_i+V_i (-\theta_i))/2V_i=0$. Later, the region $U$ has expanded to take up a volume $V_U>V_i$ which is still expanding, $\theta_U>0$, though slower than initially, $\theta_U<\theta_i$, whereas the region $O$ has shrunken to a small virialized structure with volume $V_O \ll V_U$ which is essentially static, $\theta_O \simeq 0$. In the process, the volume-averaged expansion rate has become positive: $\langle \theta \rangle = (V_U \theta_U+V_O \theta_O)/(V_U+V_O) \simeq \theta_U >0$, confirming that the global expansion has accelerated regardless of decelerating locally at each point. By taking the junction conditions between the different regions into account, the emergence of the acceleration has been verified within exact solutions of general relativity \cite{Kai:2006ws,Paranjape:2006cd}.

The current studies suggest that while the dynamical backreaction (\ref{backreaction}) can be significant, perhaps increasing the average expansion by a factor $\sim 1.3$, it may not be large enough to account for the required total increase factor $\sim 1.5$ alone \cite{Wiltshire:2007jk,Wiltshire:2007fg,Leith:2007ay,Wiltshire:2007zh,Wiltshire:2007zj,Wiltshire:2008sg}; however, it has been argued that the approximations in the current calculations may underestimate the backreaction \cite{Rasanen:2008it}. In Sect.\ \ref{selectioneffects} we propose that, due to the formation of opaque structures, the volume-averages do not necessarily capture the total effect of the inhomogeneities but the cosmic expansion may increase more rapidly along our line of sight to the observed objects than the global volume-average.

\subsection{A large local void}\label{voidmodel}

When looking at the large-scale structure of the universe, perhaps the most eye-catching feature is that the major part of the universe appears to be taken up by voids of size $\mathcal{O}(10)...\mathcal{O}(100)~{\rm Mpc}$ \cite{Hoyle:2003hc,Gott:2003pf,Tikhonov:2007di,vonBendaBeckmann:2007wt,Einasto}, nearly empty regions expanding faster than the whole universe on average. Even larger voids could exist: it has recently been suggested that the correlation between the cold spot in the CMB and the deficit of radiogalaxies in the same direction might be evidence for a void of size $\sim 300~\rm{Mpc}$ \cite{Rudnick:2007kw}; though see \cite{Smith:2008tc}. It could also be possible that we would happen to live inside such a void \cite{Moffat:1994qy,Zehavi:1998gz,Tomita:2000jj,Frith:2005et}, also known as Hubble Bubble in this context.

Clearly, if we were sitting in a special spatial location, such as inside a large local void, the global spatial averages may fail to describe our cosmological observations. In that case, an option\footnote{Another option could be to consider a different averaging scale for each redshift \cite{Mattsson:2007qp}.} is to give up the notion of a global scale factor and study an inhomogeneous metric; a particularly useful metric for describing voids is the spherically symmetric Lema\^itre-Tolman-Bondi metric, an exact solution of the Einstein equations that Lema\^itre discovered in 1933 \cite{Lemaitre:1933qe}; for a review, see \cite{Plebanski:2006sd}. During the recent years, the LTB solution has been actively exploited in studying voids and the effects of inhomogeneities on the cosmological observations \cite{Mustapha:1998jb,Celerier:1999hp,Alnes:2005rw,Zehavi:1998gz,Tomita:2000jj,Moffat:1994qy,Enqvist:2006cg,Iguchi:2001sq,Biswas:2006ub,Tanimoto:2007dq,Alexander:2007xx,GarciaBellido:2008nz,GarciaBellido:2008yq,Zibin:2008vj}.

A key notion in understanding the physical basis of the inhomogeneity induced perception of acceleration is that the cosmological observations are made along our past light cone. A way to illustrate this is to consider the directional derivative along the light cone
\begin{equation}\label{directionalderivative}
\frac{d}{dz} =  \frac{\partial x^{i}}{\partial z} \frac{\partial}{\partial x^{i}} + \frac{\partial t}{\partial z} \frac{\partial}{\partial t} \approx H_0^{-1} \left( \frac{\partial}{\partial r} - \frac{\partial}{\partial t} \right)~,
\end{equation}
where the approximation in the last step is more accurate for small distances, $r \ll H_0^{-1}$, but the sign is correct even for larger $r$. Eq.\ (\ref{directionalderivative}) tells us that negative radial variation roughly corresponds to positive time variation and that their relative contributions to the total variation w.r.t.\ the observable redshift cannot be distinguished. This is natural since by looking at a source, we simultaneously look into the past and spatially further.

Within the LTB model, the argument can be made more precise by comparing the expression of the luminosity distance in a matter dominated local void \cite{Mattsson:2007qp}
\begin{eqnarray}\label{lumiLTB}
d_L^{{\rm LTB}}(z) = H_0^{-1}(0) \bigg[ z + \left( \frac{1}{4} - \frac{H_0'(0)}{H_0^2(0)} \right) z^2 + \nonumber \\
+ \bigg( -\frac{1}{8} + \frac{1}{3} \frac{H_0'(0)}{H_0^2(0)} + 2 \frac{H_0'^2(0)}{H_0^4(0)} - \frac{1}{2} \frac{H_0''(0)}{H_0^3(0)} \bigg) z^3 + \mathcal{O} (z^4) \bigg]~,
\end{eqnarray}
where $' \equiv \partial / \partial r$, with its expression in the homogeneous and flat $\Lambda$CDM model:
\begin{equation}\label{lumiFRW}
d_L^{\Lambda{\rm CDM}}(z) = H_0^{-1} \bigg[ z + \left( \frac{1}{4} + \frac{3}{4} \Omega_{\Lambda} \right) z^2+ \left(- \frac{1}{8} - \Omega_\Lambda + \frac{9}{8} \Omega_\Lambda^2 \right) z^3 + \mathcal{O} (z^4) \bigg]~.
\end{equation}
The comparison between Eqs.\ (\ref{lumiLTB}) and (\ref{lumiFRW}) confirms that negative radial variation of the expansion rate, $H_0'(0)<0$, is needed to mimic positive cosmological constant, $\Omega_\Lambda>0$, or to induce accelerated expansion along our line of sight. In other words, for the expansion rate to increase towards us along the past light cone, the LTB expansion function $H_0(r)$ must \textit{decrease} as $r$ grows. This is exactly what an observer inside a void would see, and has been shown to account for various cosmological observations without local acceleration of the expansion \cite{Alnes:2005rw,Biswas:2006ub,Enqvist:2006cg,Alexander:2007xx,GarciaBellido:2008nz,GarciaBellido:2008yq}.

However, the explanation of the perceived cosmic acceleration as an effect induced by a local void does not come for free. Firstly, to respect the cosmological principle, we should not sit in the center of a spherically symmetric void, like fitting a local void for the observations seem to require; e.g.\ for the LTB model to remain consistent with the observed CMB dipole without a fine-tuned peculiar velocity, we should reside within a few percent of the size of the void from its center \cite{Alnes:2006pf}. In addition, even if the void were not perfectly spherical and even if we were not near the center of the void, the cosmological principle would still be, at least weakly, violated: there are less galaxies in a void and if born in a random galaxy, it is much more probably located in an environment with lots of galaxies than in a relatively empty void. Another issue is that the required size of the local void seems to be a few times bigger than the typical size of the observed large voids. Moreover, the existence of such a void has not been observationally confirmed, although, due to most of the matter in the universe being dark, it may not be easy to disprove it either. In any case, without a natural explanation for these issues, the scenario with a large local void -- if taken literally -- does not appear likely. Instead, if considered to describe observations averaged over the celestial sphere, the LTB solution may still have relevance as an effective description for light propagation through several smaller voids.

\subsection{Selection effects due to opaque structures}\label{selectioneffects}

Even if we were not located \emph{inside} a large local void, the observed cosmic network of smaller voids can still significantly affect the observable distance-redshift relations. Firstly, as discussed in Sect.\ \ref{noijannuoli}, the spatial variations in the expansion rate can increase the volume-averaged global expansion rate. In addition, a complementary aspect we want to point out in this work is that the opacity of the cosmic structures can bias the observations, such that the light we see has traveled mostly in freely expanding voids at late times, and thereby affect the observed distance-redshift relations, independently from the dynamical backreaction. Although it may be difficult to distinguish the observational consequences of the selection effects and the backreaction, they are in principle distinct mechanisms: one could have large opaque structures that induce selection effects but give negligible backreaction or, on the other hand, perfectly transparent but highly nonlinear structures that produce backreaction but no selection effects. Naturally, in the real universe, both mechanisms may have significant effects on the observations. Mathematically, the difference between the purely dynamical backreaction and the biased light propagation in voids can be seen as the difference between the volume-averaged expansion rate and the average expansion rate along our line of sight.

The selection effects in the observations of cosmic light can be justified by the fact that the emptier a region in space, the easier it is to see an object through it. The bias can be further categorized roughly as two different cases: 1) an opaque structure on the foreground simply screens the further object behind it, or 2) the further object is seen but, due to foreground contamination, not clearly enough that the object would contribute to the high quality data. The first case is perhaps relatively rare, since only the nearest galaxies, including the Milky Way, cover a notable fraction of the celestial sphere. Instead, the second case may be more relevant as, due to their long distance, the cosmological objects tend to be dim, which means that they can be harder to detect if there are bright foreground objects nearby on the sky. In addition to the brightness of the foreground, the contamination can arise from gravitational lensing. An important implication of the contamination is that the effective area screened by the foreground objects may be significantly larger than their physical cross-section. As an example, large type Ia supernova surveys typically cover only $\sim 10^{-4}$ part of the sky \cite{Pain:2002wj}. It thus appears conceivable that the well-observed supernovae may consist more of objects with less than average intervening material on their foreground.

In addition to the kind of bias discussed above, there may be other kind of selection effects, arising from the fact that the dimmer the object the harder it is to detect. It appears non-trivial how this would affect the observed distance-redshift relations, since on one hand it could favor objects with higher luminosity but on the other hand disfavor objects with high redshift. In any case, for the statistically most relevant brighter supernovae at lower redshifts, we would expect this effect to be less significant. Furthermore, if the type Ia supernovae would be perfect standard candles, the bias towards higher intrinsic luminosity should not exist at all. In addition to selection effects, there could also be other systematic effects related to observational techniques \cite{Tonry:2004}. Here we simply assume that such effects do not play a significant role.

Unfortunately, due to the complexity of the cosmological structures, it appears extremely difficult to derive quantitative estimates for the significance of the selection effects. A possible way to obtain such estimates could be a computer simulation where one studies the difference of the average expansion rate along our line of sight relative to the volume-averaged global expansion rate. Altogether, we can currently neither vindicate nor rule out the importance of cosmic selection effects from opaque structures.

\subsection{Total effect of the inhomogeneities}\label{TotalEffect}

As we have discussed in Sects.\ \ref{noijannuoli}--\ref{selectioneffects}, it appears very difficult to derive from theory a realistic inhomogeneous model that would take into account all the relevant effects of the inhomogeneities. Instead of dealing with these difficulties, we employ a phenomenological approach based on physically justified assumptions in Sect.\ \ref{themodel}. The model contains a transition redshift to the void-dominated era as a new free parameter and the applied mathematical framework is provided by a generalized Dyer-Roeder method, introduced in Sect.\ \ref{dyerroeder}. In this approach, the observable or optical distance-redshift relations are determined by the average properties of the space along our line of sights to the objects, instead of global volume-averages. The aim of the model is to provide only a simple description of what the overall effect of the cosmic inhomogeneities on the observed distance-redshift relations may be. Although based on assumptions, in the end, one can use the naturalness of the required value for the new parameter to judge the viability of the model, e.g.\ compared to the standard FRW cosmology with $\Lambda$ as the additional phenomenological parameter.

Perhaps the first to discuss light propagation in the clumpy universe was Zel'dovich in 1963 \cite{Zeldovich}. Before that, the idealized FRW geometry had been used as the basis in deriving theoretical predictions for the observable distance-redshift relations. Later in the 60's, Zel'dovich's idea was elaborated at least by Bertotti \cite{Bertotti}, Gunn \cite{Gunn1} and Kantowski \cite{Kantowski1}. However, it was the works by Dyer and Roeder in the 70's that made the idea more famous \cite{DyerRoeder1,DyerRoeder2}; consequently, the distance-redshift relations allowing for the effects of clumping of matter on light are usually referred to as the Dyer-Roeder relations.

It seems that the interest towards the effects of clumpiness on light propagation suffered a hiccup after 1976, when Weinberg argued against the use of the Dyer-Roeder version of the distance-redshift relations in place of the conventional FRW ones \cite{Weinberg1}. He proposed that the effect of light propagating in the empty intergalactic space would, on average, be canceled by an equal but opposite effect due to gravitational lensing caused by the matter clumps. However, Weinberg's argument has later been challenged \cite{Ellis:1998ha}. In addition, we want to point out that the original Dyer-Roeder relations do not take two crucial physical properties of the universe into account: the inhomogeneities in the \emph{expansion} rate and the \emph{growth} of the inhomogeneities. Indeed, in Sect.\ \ref{themodel} we demonstrate that the evolving inhomogeneous expansion can induce much stronger corrections to the observable distance-redshift relations than the mere clumpiness of matter in the original formulation of Dyer and Roeder -- and hence unlikely to get counterbalanced by the gravitational lensing.

\subsection{The original Dyer-Roeder method}\label{originalDR}

Before presenting the generalized version of the Dyer-Roeder method in Sect.\ \ref{dyerroeder}, we introduce the original formulation here for reference and to justify the modifications. For more discussion on the method, see Refs.\ \cite{DyerRoeder1,DyerRoeder2,Linder1,Kantowski:1995bd,Kantowski:1998ju,Kantowski:2003yt}.

A good starting point for quantitative considerations of light propagation in the inhomogeneous universe is the exact Sachs optical scalar equations \cite{Sachs}. They correspond to the scalar part of the Einstein tensor equation on null geodesics. The first one reads as
\begin{equation}\label{sachs1}
\frac{\partial \tilde{\theta}}{\partial \lambda} + \frac{1}{2} \tilde{\theta}^2 + \tilde{\sigma}^2 + R_{\mu \nu} k^\mu k^\nu = 0~,
\end{equation}
where $\tilde{\theta} \equiv V^{-1} \partial V / \partial \lambda$ is the expansion of the bundle of light rays taking up a volume $V$; $\lambda$ is the affine parameter along the null geodesics $x^\mu(\lambda)$ that gives the energy of the photons as measured by an observer with four-velocity $u^\mu$ via $E(\lambda)=- g_{\mu \nu} u^\mu \partial x^\nu/ \partial \lambda$; $R_{\mu \nu}$ is the Ricci curvature tensor, whereas $\tilde{\sigma}^2$ represents optical shear and $k^\mu \equiv \partial x^\mu/ \partial \lambda$ is the wave-vector of the light rays. The other Sachs equations are trivially satisfied when the optical shear is neglected as in the Dyer-Roeder approach and which is also an in-built assumption of the standard FRW description.

To rewrite Eq.\ (\ref{sachs1}) in terms of observables, we use $\tilde{\theta} = A^{-1} \partial A / \partial \lambda $ where $A$ is the cross-sectional area of the light bundle and collect the derivatives into one term as
\begin{equation}\label{sachs1b}
\frac{\partial \tilde{\theta}}{\partial \lambda} + \frac{1}{2} \tilde{\theta}^2 = \frac{2}{\sqrt{A}} \frac{\partial^2 \sqrt{A} }{\partial \lambda^2}~,
\end{equation}
to obtain
\begin{equation}\label{sachs1c}
\frac{2}{d_A} \frac{\partial^2}{\partial \lambda^2} {d^{\phantom{A}}_{A}} + 8 \pi G T_{\mu \nu} k^\mu k^\nu = 0~,
\end{equation}
where the Einstein equation has been employed in the form $R_{\mu \nu} = 8 \pi G T_{\mu \nu} + g_{\mu \nu} R/2$ with the facts that $k^{\mu}$ is null, $g_{\mu \nu} k^{\mu} k^{\nu} = 0$, and that the angular diameter distance is proportional to the square root of the cross sectional area of the light bundle, $d_A \propto \sqrt{A}$.

By using ideal fluid to describe the energy content of the universe, we have the energy tensor $T^{\mu \nu} = (\rho+p) u^\mu u^\nu + p g^{\mu \nu}$ where $u^\mu = \delta^\mu_0$ in the comoving coordinates and $\rho=\rho_m + \rho_r$ stands for the energy density of matter plus radiation and $p=p_r$ is the radiation pressure. Hence, with the help of $g_{\mu \nu} u^\mu k^\nu = -E(\lambda)$, Eq.\ (\ref{sachs1c}) yields
\begin{equation}\label{sachsi1}
\frac{\partial^2}{\partial \lambda^2} {d^{\phantom{A}}_{A}}(\lambda)  + 4 \pi G (\rho+p) E_0^2 (1+z)^2 d_A (\lambda)= 0~,
\end{equation}
where we have used the definition of the redshift, $E(\lambda)/E_0 \equiv 1+z$, with $E_0$ as the observed energy of the photons today. We still want the affine curve parameter $\lambda$ in terms of the non-affine but observable redshift $z$:
\begin{equation}\label{redshiftaffine}
\frac{\partial}{\partial \lambda} = \frac{\partial t}{\partial \lambda} \frac{\partial z}{\partial t} \frac{\partial}{\partial z} = E(\lambda) \left[-(1+z)H(z)\right] \frac{\partial}{\partial z} = -E_0 (1+z)^2 H(z) \frac{\partial}{\partial z}~,
\end{equation}
where $H(z)$ is the Hubble expansion along the path of the light and it has been assumed that the relation between the auxiliary scale factor and the redshift, $a(0)/a(z) = 1+z$, is a good approximation and that the proper time $t$ is independent of the spatial location (which is only possible if the vorticity of matter vanishes). Finally, by using Eq.\ (\ref{redshiftaffine}) we can write Eq.\ (\ref{sachsi1}) as
\begin{equation}\label{sachsi1b}
H(z) \frac{\partial}{\partial z} \left[ (1+z)^2 H(z) \frac{\partial}{\partial z} d_A(z) \right] + 4 \pi G \left[ \rho(z) + p(z) \right] d_A(z) = 0~,
\end{equation}
where the dependence on the energy of the photons $E_0$ canceled out.

By using the general relation between the angular and luminosity distances,
\begin{equation}\label{luminosityangular}
d_L(z) = (1+z)^2 d_A(z)~,
\end{equation}
which holds for geodesic light in any spacetime as proved by Etherington in 1933 \cite{Etherington,Ellis,Plebanski:2006sd}, Eq.\ (\ref{sachsi1b}) can be transformed into the corresponding equation for the luminosity distance:
\begin{equation}\label{sachsi1c}
(1+z)^2 H(z) \frac{\partial}{\partial z} \left[ (1+z)^2 H(z) \frac{\partial}{\partial z} \frac{d_L(z)}{(1+z)^2} \right] + 4 \pi G \left[\rho(z) + p(z) \right] d_L(z) = 0~.
\end{equation}

The essential point Dyer and Roeder made was that, due to Eqs.\ (\ref{sachsi1b}) and (\ref{sachsi1c}) determining the optical properties of the universe, the matter density $\rho_m$ should be interpreted as the average density of the intergalactic regions where the observed light has traveled. To describe this, they introduced a constant \emph{smoothness parameter} $\alpha$, representing the fraction of the matter density along our line of sight $\rho_m$ to the volume-averaged matter density $\bar{\rho}_m$ of the whole universe in the FRW description. That is, $\rho_m \equiv \alpha \bar{\rho}_m$ in Eqs.\ (\ref{sachsi1b}) and (\ref{sachsi1c}), with the value $\alpha=0$ corresponding to a late universe where all the matter is concentrated into opaque clumps and the value $\alpha=1$ to the smooth early universe. Clearly, $\alpha$ can then be something between these two extremes.

Eqs.\ (\ref{sachsi1b}) and (\ref{sachsi1c}) are second order differential equations for $d_A(z)$ and $d_L(z)$ with the initial conditions: $d_A(0)=0$, ${d_A}'(0) = H_0^{-1}$ and $d_L(0)=0$, ${d_L}'(0) = H_0^{-1}$, where $' \equiv \partial/\partial z$. Thus, we can easily write the solution to Eq.\ (\ref{sachsi1c}) as a power series in $z$:
\begin{equation}\label{lumiDR}
d_L^{{\rm DR}}(z) = H_0^{-1} \bigg[ z +  \frac{1}{4} z^2 + \left( - \frac{1}{8} + \frac{1- \alpha}{4} \right) z^3 + \mathcal{O} (z^4) \bigg]~,
\end{equation}
where it has been assumed that the Hubble expansion has the flat FRW form with pressureless matter as the only source, $H(z) = H_0 (1+z)^{3/2}$, since the radiation is negligible at small redshifts where the solution (\ref{lumiDR}) would only hold anyway.

By comparing the Dyer-Roeder luminosity distance (\ref{lumiDR}) with its counterpart in the homogeneous $\Lambda$CDM model (\ref{lumiFRW}), we see that there is no way $\alpha$ could mimic $\Lambda$, as it does not even appear in the all-important second-order term. In fact, the effect of clumping via $\alpha<1$ seems to effectively \emph{increase the deceleration} of the expansion in concordance with the original results of Dyer and Roeder \cite{DyerRoeder1,DyerRoeder2}.

However, on physical grounds, one could expect just the opposite result: due to the local gravitational attraction of matter, a shearless and irrotational region of space with low mass density should decelerate \emph{less} than a denser one. This can also be seen from the Raychaudhuri equation, an exact scalar part of the Einstein equation \cite{Plebanski:2006sd}:
\begin{equation}\label{Raychaudhuri}
- \frac{1}{V^{1/3}} \frac{d^2 }{d \tau^2} V^{1/3} = \frac{4 \pi G}{3} \rho_m + \frac{1}{3} (\sigma^2-\omega^2)~,
\end{equation}
where $\sigma^2$ and $\omega^2$ represent the shear and vorticity of matter respectively, $\tau$ is the proper time of the matter particles and $V$ is a local volume element, making $V^{1/3}$ a generalized scale factor. As expected, in a region far enough from the high-density filaments so that $\sigma^2-\omega^2 \ll 4 \pi G \rho_m$, the deceleration or the left hand side of Eq.\ (\ref{Raychaudhuri}) is the smaller the lower the local matter density $\rho_m$.

We argue that these physically reasonable conclusions do not show up in the solution (\ref{lumiDR}) due to the incomplete description of the inhomogeneities in the original Dyer-Roeder method. Firstly, matter density $\rho_m$ is not the only quantity with inhomogeneous distribution, but the local Hubble expansion along our line of sight $H(z)$ can differ from the FRW value as well. In fact, the effect of the expansion rate on the distance-redshift relations is more dominant than the effect of the matter extent, because the expansion is mainly responsible for the redshift. This has been demonstrated within the LTB model in Ref.\ \cite{Enqvist:2006cg} and is manifest already in the distance-redshift relations of the FRW models:
\begin{equation}\label{lumiFRWgeneral}
\bar{d}_L(z) = (1+z) \frac{H_0^{-1}}{\sqrt{|1-\Omega_0|}} S_k
\left[ \sqrt{|1-\Omega_0|} H_0 \int_{0}^{z} \frac{d\tilde{z}}{H(\tilde{z})} \right] ~,
\end{equation}
with only a relatively weak dependence on the overall density $\Omega_0$ via
\begin{equation} S_k(x) = \left\{\begin{array}{cl}
\sinh (x) & $if $ \Omega_0 < 1 \\
x         & $if $ \Omega_0 = 1 \\
\sin (x)  & $if $ \Omega_0 > 1
\end{array}\right . \label{gensin} , \end{equation}
whereas the expansion integrated over the past light cone, $\int_{0}^{z} H^{-1}(\tilde{z}) d\tilde{z}$, yields the main contribution to the relation (\ref{lumiFRWgeneral}).

Altogether, we propose that in order to make the method physically more correct, the Hubble expansion $H(z)$ in Eqs.\ (\ref{sachsi1b}) and (\ref{sachsi1c}) should be interpreted as the average expansion rate of the regions where the detectable light travels. For this, we introduce another parameter $\beta$, the ratio of the expansion rate along our line of sight $H(z)$ to the FRW value $\bar{H}(z)$, so that $H(z) \equiv \beta \bar{H}(z)$ in Eqs.\ (\ref{sachsi1b}) and (\ref{sachsi1c}). Another crucial ingredient missing from the original formulation is that structures evolve in the universe; that is, both of the parameters should depend on the redshift: $\alpha(z)$ and $\beta(z)$. The redshift-dependence of $\alpha$ was in fact already proposed by Linder in 1988 \cite{Linder1}, but perhaps because the corrections in the original Dyer-Roeder method are so weak (see Eq.\ (\ref{lumiDR})), the redshift dependence of $\alpha$ has won very little attention in the literature \cite{Santos:2008tz}. Instead, as we demonstrate in Sect.\ \ref{dyerroeder}, it is only after taking into account \emph{both} the parameter $\beta$ \emph{and} its dependence on the redshift that the results start to go more hand in hand with intuition.

\section{Optically altered universe due to nonlinear structures}\label{themodel}

In this section, we introduce the principal model of this work, where light propagation in the real clumpy universe is described phenomenologically by a generalized Dyer-Roeder method. Within this approach, the universe can still be homogeneous on large scales, but the nonlinear inhomogeneities on smaller scales render its optical properties different. As discussed in Sect.\ \ref{classification}, the optical overall effect of the inhomogeneities can result from: 1) biased light propagation through voids that expand faster than the average, 2) global increase in the volume-averaged expansion due to spatial variations in the expansion rate, 3) our possible special spatial location in the universe, or any combination of the three. There can also be effects that are not reducible to any of these; e.g.\ a clock-rate difference between underdense and overdense regions discussed in Refs.\ \cite{Wiltshire:2007fg,Wiltshire:2007jk,Leith:2007ay,Wiltshire:2007zh,Wiltshire:2007zj,Wiltshire:2008sg}.

We proceed by summarizing the physical foundations of the model qualitatively in Sect.\ \ref{physics} and leave the discussion of its quantitative properties to Sects.\ \ref{dyerroeder} -- \ref{results}.

\subsection{The physical foundations of the model}\label{physics}

The assumptions that form the physical foundations of the model, can be summarized in three main points:
\begin{enumerate}
\item The older the universe, the clumpier it is and hence the emptier and larger the voids.
\item The light from the well-observed objects travels mostly in the voids.
\item The emptier and larger the voids, the faster they expand relative to the average.
\end{enumerate}
The first point is the standard notion of structure formation: initially overdense regions accrete more material via gravitational attraction and thus become denser and denser at the expense of the initially underdense regions, which thereby become emptier. Some observational evidence for the emptiness of voids has been found in galaxy surveys that show there to be typically $\lesssim 1\%$ galaxies near the center of the voids relative to the average galaxy density \cite{Hoyle:2003hc}. The second assumption was explained in Sect.\ \ref{selectioneffects}: the emptier a region the easier it is to see an object through it. The point three is an integrated effect of Eq.\ (\ref{Raychaudhuri}): denser regions have more mass and consequently stronger gravitational attraction to slow down the local expansion.

The combination of these properties suggests a physical explanation for how the nonlinear structure formation can be perceived as accelerated expansion without dark energy: as the detectable light traverses emptier and emptier regions, the expansion rate along the path of the photons increases relative to the average, causing accelerated expansion along our line of sight without local acceleration. More specifically, we present the following conjecture: due to the formation of voids and opaque structures, the average matter density on the path of the light from the well-observed objects changes from $\Omega_M \simeq 1$ in the homogeneous early universe to $\Omega_M \simeq 0$ in the clumpy late universe, so that the average expansion rate increases along our line of sight from EdS expansion $Ht \simeq 2/3$ at high redshifts to free expansion $Ht \simeq 1$ at low redshifts.

\emph{Optically}, the proposed phenomenon bears close resemblance to the local Hubble Bubble, discussed in Sect.\ \ref{voidmodel}; see in particular Eqs.\ (\ref{lumiLTB}), (\ref{lumiFRW}) and the paragraph thereafter. However, unlike in the models with a large void in the local matter distribution, it is our special location in time, not in space, that is here crucial for the effect, so the cosmological principle is \emph{not} violated. On the contrary, since every observer in the late universe sees the history of the structure formation, the phenomenon should look essentially the same no matter at which point the observer sits in the universe\footnote{This, at least, seems to distinguish the light propagation in freely expanding voids as the cause for the acceleration from Wiltshire's proposal whereby it is crucial that the observer sits in a static or gravitationally bound location, such as inside a galaxy, to perceive the inhomogeneity induced acceleration \cite{Wiltshire:2007jk}.}.

\subsection{The generalized Dyer-Roeder method}\label{dyerroeder}

In Sect.\ \ref{originalDR}, we reviewed a quantitative method to allow for the optical effects of the clumpiness of matter in the universe, following the reasoning of Dyer and Roeder \cite{DyerRoeder1,DyerRoeder2}. The basic idea of this method is to employ the average density along the paths from the objects to the observer in calculating the distance-redshift relations for those objects, instead of using an average density of the whole universe like in the FRW description. Albeit a promising starting point, we argued that the original method falls short of capturing two crucial physical phenomena: inhomogeneities in the expansion rate and the growth of the nonlinear structures. To take these properties into account, we present a generalized version of the original Dyer-Roeder method.

Following the argumentation of Sect.\ \ref{originalDR}, we take into account both the inhomogeneities in the matter distribution and in the expansion rate via the redshift-dependent parameters $\alpha(z)$ and $\beta(z)$. Continuing to neglect the vorticity of matter, we still have a coordinate system where the matter four-velocity is just $u^\mu = \delta^\mu_0$. Thus, we can simply substitute $\rho_m(z) \rightarrow \alpha(z) \bar{\rho}_m(z)$ and $H(z) \rightarrow \beta(z) \bar{H}(z)$ in Eqs.\ (\ref{sachsi1b}) and (\ref{sachsi1c}) to obtain
\begin{equation}\label{dAmodifiedDR}
\beta(z) \bar{H}(z) \frac{\partial}{\partial z} \left[ (1+z)^2 \beta(z) \bar{H}(z) \frac{\partial}{\partial z} d_A(z) \right] + 4 \pi G \left[ \alpha(z) \bar{\rho}_m(z) + \frac{4}{3} \rho_r(z) \right] d_A(z) = 0
\end{equation}
\begin{equation}\label{dLmodifiedDR}
(1+z)^2 \beta(z) \bar{H}(z) \frac{\partial}{\partial z} \left[ (1+z)^2 \beta(z) \bar{H}(z) \frac{\partial}{\partial z} \frac{d_L(z)}{(1+z)^2} \right] + 4 \pi G \left[ \alpha(z) \bar{\rho}_m(z) + \frac{4}{3} \rho_r(z) \right] d_L(z) = 0~,
\end{equation}
where the symbols with an overbar represent the auxiliary FRW quantities and we assume that radiation does not cluster: $\rho_r(z)=\bar{\rho}_r(z)$. The initial conditions for the differential equations (\ref{dAmodifiedDR}) and (\ref{dLmodifiedDR}) are: $d_x(0)=0$, ${d_x}'(0) = H_0^{-1}$, where $x \in \{A,L\}$.

To work out how the inhomogeneities can modify the luminosity distance in the generalized Dyer-Roeder method, consider the solution to Eq.\ (\ref{dLmodifiedDR}) as a power series in $z$:
\begin{equation}\label{lumiDRmodified}
H_0d_L(z) = \bigg[ z + \left(1 - \frac{\beta'(0)}{2\beta(0)} \right) z^2 + \left( \frac{\beta'^2(0)}{3\beta^2(0)} - \frac{\beta'(0)}{3\beta(0)} - \frac{\beta''(0)}{6\beta(0)} - \frac{\alpha(0)}{4\beta^2(0)} \right) z^3 + \mathcal{O}(z^4)\bigg],
\end{equation}
where radiation has again been neglected. Here the nonlinear structures, encapsulated in the functions $\alpha(z)$ and $\beta(z)$, have much more notable role than in the original Dyer-Roeder method in Eq.\ (\ref{lumiDR}). In addition, as can be seen by comparing the solution (\ref{lumiDRmodified}) to the luminosity distance in the flat $\Lambda$CDM model (\ref{lumiFRW}), the effect of the inhomogeneities is here importantly in the correct direction, mimicking $\Lambda$ with $\beta'(0)<0$.

We emphasize that the solutions $d_A(z)$ and $d_L(z)$ to Eqs.\ (\ref{dAmodifiedDR}) and (\ref{dLmodifiedDR}) do not represent mathematically exact solutions of general relativity. Instead, their aim is to capture the most essential effects of the inhomogeneities on the observable properties of light that the simplified FRW description fails to describe. The crudest approximations in Eqs.\ (\ref{dAmodifiedDR}) and (\ref{dLmodifiedDR}) correspond to employing a different FRW result for each redshift. Even approximations of this kind have proven to be accurate in calculating the distance-redshift relations \cite{Mattsson:2007qp}. Nevertheless, in a more thorough treatment, e.g.\ the utilized FRW relation $a(0)/a(z) = 1+z$ should be replaced with a corrected inhomogeneous result; see \cite{Rasanen:2008be}.

To calculate the observable distance-redshift relations $d_A(z)$ and $d_L(z)$ from Eqs.\ (\ref{dAmodifiedDR}) and (\ref{dLmodifiedDR}), we need the functions $\alpha(z)$ and $\beta(z)$. An entirely phenomenological approach would be to determine them purely from the cosmological data analysis. In contrast, an ideal approach would be to derive $\alpha(z)$ and $\beta(z)$ from the full Einstein equations starting with some initial perturbations in the early universe. In this work, we employ an analytic approach which falls somewhere between these two extremes: based on theoretical and observational reasoning, we deduce a form for the functions $\alpha(z)$ and $\beta(z)$ containing a single phenomenological parameter that, given the approximations, would best correspond to the structure formation of the real universe; this is done in Sect.\ \ref{derivation}. In Sect.\ \ref{observations}, we then fit the leftover parameter to various cosmological data.

\subsection{A phenomenological description of the nonlinear structure formation}\label{derivation}

As boundary conditions given on a 3-dimensional hypersurface, the expansion rate $H$ and the matter distribution $\rho_m$ are independent (apart from some constraints). In theory, one could thus give them almost any profile e.g.\ along our past light cone; that is, choose $\alpha(z)$ and $\beta(z)$ in Eqs.\ (\ref{dAmodifiedDR}) and (\ref{dLmodifiedDR}) arbitrarily. However, as suggested by the nearly uniform temperature of the CMB radiation, it is likely that the early universe has been very close to homogeneous. Thus, we only consider inhomogeneities that grow with time, which sets a constraint between the functions $H(z)$ and $\rho_m(z)$, or equivalently between $\alpha(z)$ and $\beta(z)$, reducing the number of independent functions to one.

Some properties of $\alpha(z)$ and $\beta(z)$ can be deduced without resorting to a specific model. Firstly, according to Eq.\ (\ref{Raychaudhuri}), initial underdensities evolve into regions that eventually begin to expand faster than the global average. As the detectable light is assumed to propagate mostly in these regions, we have $\alpha<1$, hence implying that $\beta>1$. The requirement of positive mass density sets $\alpha \geq 0$, while Eq.\ (\ref{Raychaudhuri}) implies the upper bound $\beta \leq 3/2$. For an explicit constraint between $\alpha(z)$ and $\beta(z)$, a specific model is required.

The question for which we seek an answer is then:\ how much faster does a void with known matter density $\rho_m < \bar{\rho}_m$ expand than the FRW value $\bar{H}$, i.e.\ given $0 \leq \alpha<1$, what is $1 < \beta \leq 3/2$? We estimate this constraint by employing a different FRW result to each patch with different density and expansion rate. Given that we use the proper time of dust as our time coordinate, the constraint can be obtained by requiring a constant value for the age of the universe $t_0$ between the different regions. This sets the matter density and expansion rate equal at $t=0$ for every patch, i.e.\ the condition that the universe was initially homogeneous. We do not, however, employ this method to model the dynamics of the universe as, e.g.\ due to vanishing vorticity and ignored mass flows, it cannot realistically describe the evolution of the structures at least when virialization becomes important.

As a starting point, we take the functions $\alpha(z)$ and $\beta(z)$ to measure deviation from a flat FRW model with matter and radiation. Thus the quantities with an overbar in Eqs.\ (\ref{dAmodifiedDR})--(\ref{dLmodifiedDR}) obey the relations
\begin{equation}\label{dynamicalRho}
4 \pi G \bar{\rho}_m(z) = \frac{3}{2} \bar{H}_0^2 \bar{\Omega}_m (1+z)^3
\end{equation}
\begin{equation}\label{dynamicalHubble}
\bar{H}(z) = \bar{H}_0 (1+z)^{3/2} \sqrt{1+ z \bar{\Omega}_r}~,
\end{equation}
where $\bar{\Omega}_m \equiv 8 \pi G \bar{\rho}_m(0)/(3 \bar{H}_0^2)$ and $\bar{\Omega}_r \equiv 8 \pi G \bar{\rho}_r(0)/(3 \bar{H}_0^2)$ are the density parameters of the auxiliary FRW model (with the constraint $\bar{\Omega}_m+\bar{\Omega}_r=1$) and $\bar{H}_0$ would be interpreted as the average global expansion rate in the FRW description. However, if the dynamical backreaction (\ref{backreaction}) is significant, in the nonlinear late universe $\bar{\rho}_m(z)$ and $\bar{H}(z)$ are just auxiliary quantities with no obvious physical interpretation.

To estimate the constraint between $\alpha(z)$ and $\beta(z)$, we first replace them by the auxiliary functions $\tilde{H}_0(z)$ and $\tilde{\Omega}_m(z)$, defined through the equations:
\begin{equation}\label{alfaFRW}
\alpha(z) = \frac{\tilde{H}_0^2(z) \tilde{\Omega}_m(z)}{\bar{H}_0^2 \bar{\Omega}_m}~
\end{equation}
\begin{equation}\label{betaFRW}
\beta(z) = \frac{\tilde{H}_0(z)}{\bar{H}_0}\sqrt{\frac{1+ z \tilde{\Omega}_m(z)+z (z+2) \bar{\Omega}_r}{(1+z)(1+ z \bar{\Omega}_r)}}~.
\end{equation}
With the definitions (\ref{alfaFRW})--(\ref{betaFRW}), the line-of-sight quantities $\rho_m(z)$ and $H(z)$ take the form
\begin{equation}\label{RhozFRW}
4 \pi G \rho_m(z) = 4 \pi G \alpha(z) \bar{\rho}_m(z) = \frac{3}{2} \tilde{H}_0^2(z) \tilde{\Omega}_m(z) (1+z)^3~
\end{equation}
\begin{equation}\label{HubblezFRW}
H(z) = \beta(z) \bar{H}(z)  = \tilde{H}_0(z) (1+z) \sqrt{1+ z \tilde{\Omega}_m(z)+z (z+2) \bar{\Omega}_r}~,
\end{equation}
where, by construction, the deviation from homogeneity appears as the redshift dependence of $\tilde{H}_0(z)$ and $\tilde{\Omega}_m(z)$ which in the limit $z \gg 1$ reduce to the constants $\bar{H}_0$ and $\bar{\Omega}_m$. Apart from this limit, there is no straightforward physical interpretation for $\tilde{H}_0(z)$ and $\tilde{\Omega}_m(z)$.

The next step is to obtain the constraint between $\tilde{H}_0(z)$ and $\tilde{\Omega}_m(z)$ by requiring that the age of the universe is the same between the different FRW patches. The relation between $\tilde{H}_0(z)$ and $\tilde{\Omega}_m(z)$ then determines a constraint between $\alpha(z)$ and $\beta(z)$ through Eqs.\ (\ref{alfaFRW}) and (\ref{betaFRW}). We are only applying the FRW results to voids, so $\Omega_m \leq 1$. In this case, the integration of the Friedman equation yields the age of the FRW dust universe:
\begin{equation}\label{FRWage}
t_{0} = \frac{\sqrt{1-\Omega_m} - \Omega_m {\rm{arsinh}}\sqrt{\frac{1-\Omega_m}{\Omega_m}} }{H_0 (1-\Omega_m)^{3/2}}~.
\end{equation}
We use the redshift to label the different patches on the present-day spatial hypersurface, so $\Omega_m$ and $H_0$ in Eq.\ (\ref{FRWage}) become functions of $z$. The requirement for the age in Eq.\ (\ref{FRWage}) to be equal between the different patches then yields $H_0$ in terms of $\Omega_m$:
\begin{equation}\label{constraintFRW}
\tilde{H}_0(z) = t_0^{-1} \left[ \frac{\sqrt{1-\tilde{\Omega}_m(z)} - \tilde{\Omega}_m(z)
{\rm{arsinh}} \sqrt{\frac{1-\tilde{\Omega}_m(z)}{\tilde{\Omega}_m(z)}}
}{(1-\tilde{\Omega}_m(z))^{3/2}} \right] \simeq t_0^{-1} \left[ 1 -\frac{1}{3} \sqrt{\tilde{\Omega}_m(z)}  \right]~,
\end{equation}
where $t_0={\rm{constant}}$ is the age of the universe and in the last step we have employed an approximation which is accurate ($|{\rm error}|$ $<1.5\%$) in the interval $0 \leq \tilde{\Omega}_m(z) \leq 1$.

Using Eqs.\ (\ref{alfaFRW}), (\ref{betaFRW}) and (\ref{constraintFRW}), we can give $\tilde{\Omega}_m(z)$ and $\tilde{H}_0(z)$ in terms of $\alpha(z)$,
\begin{equation}\label{Omegaalpha}
\tilde{\Omega}_m(z) =  \frac{9}{4} \left( 1 - \sqrt{ 1- \frac{8}{9} \sqrt{\bar{\Omega}_m \alpha(z)} } \right)^2
\end{equation}
\begin{equation}\label{Halpha}
\tilde{H}_0(z) =  \frac{t_0^{-1}}{2} \left( 1 + \sqrt{ 1- \frac{8}{9} \sqrt{\bar{\Omega}_m \alpha(z)} } \right) ~,
\end{equation}
so the line-of-sight quantities (\ref{RhozFRW}) and (\ref{HubblezFRW}) become
\begin{equation}\label{RhozFRW2}
4 \pi G \rho_m(z) = \frac{2}{3} t_0^{-2} \bar{\Omega}_m \alpha(z) (1+z)^3
\end{equation}
\begin{equation*}
H(z) = \beta(z) \bar{H}(z) = \frac{t_0^{-1}}{2} (1+z) \left( 1 + \sqrt{ 1- \frac{8}{9} \sqrt{\bar{\Omega}_m \alpha(z)} } \right) \times \\
\end{equation*}
\begin{equation}\label{HubblezFRW2}
\sqrt{1+ \frac{9z}{4} \left( 1 - \sqrt{ 1- \frac{8}{9} \sqrt{\bar{\Omega}_m \alpha(z)} } \right)^2 +z (z+2) (1- \bar{\Omega}_m)}~.
\end{equation}

We impose two observationally motivated boundary values for the function $\alpha(z)$. Firstly, soon after decoupling the universe was still close to homogeneous, so the difference between the line-of-sight and globally averaged matter densities must have been negligible, corresponding to $\rho_m \rightarrow \bar{\rho}_m$ or $\alpha (1100) \simeq 1$. At times earlier than $z=1100$, the distinction between the line-of-sight and volume averaged quantities does not make sense, since we cannot see beyond the opaque plasma wall formed by the last scattering surface. The second boundary value follows from the assumption that the regions where the detectable light from the well-observed objects travels today are nearly empty: $\rho_m(0) \ll \bar{\rho}_m(0)$ or $\alpha(0) \simeq 0$. Altogether, we then have $\alpha(0) \simeq 0$ and $\alpha(1100) \simeq 1$.

With the endpoints $\alpha(0) \simeq 0$ and $\alpha(1100) \simeq 1$ of the function $\alpha(z)$ fixed, we still need its behavior between these points.\ Once initiated, the structures can only grow in a universe with no dark energy, so a logical starting point is that $\alpha(z)$ must be a monotonically increasing function of redshift. We choose the exact behavior of the function based on physical understanding about the structure formation.

A typical behavior of a small overdensity on a given scale $L$ is that initially the excess decrease in its expansion rate is small, with linear proportionality to the initial conditions; but eventually, around a redshift $z_0(L)$, the gravity-driven cumulative increase in the clustering speed of the constituents collapses the system rapidly until it virializes due to pressure gradients and the conservation of angular momentum \cite{Peacock,Springel:2005nw}.

A function that encapsulates these qualitative features is the exponential, $\alpha(z) = \alpha_0 + \sum_i \alpha_i e^{-z/z_i(L_i)}$, summed over different scales. For simplicity, we consider only a single redshift $z_0$ when the voids start to dominate at the most significant scale. Taking into account the above deduced boundary values, we choose
\begin{equation}\label{OmegaParameterized}
\tilde{\Omega}_m(z) =  \left( 1 - e^{-z/z_0}\right)^2 ~,
\end{equation}
or
\begin{equation}\label{DensityAlongOurLineOfSight}
\alpha(z) =   \frac{1}{4} \left(2 + e^{-z/z_0} \right)^2 \left(1 - e^{-z/z_0} \right)^2 ~.
\end{equation}
Inserting Eq.\ (\ref{DensityAlongOurLineOfSight}) in Eq.\ (\ref{Halpha}), we also have
\begin{equation}\label{ExpansionAlongOurLineOfSight}
\tilde{H}_0(z) t_0 \equiv h(z) = \frac{1}{3}\left(2 + e^{-z/z_0}\right)~.
\end{equation}
In the end, we leave the transition redshift $z_0$ as a free parameter. A more sophisticated numerical estimation of the function $\alpha(z)$ will be performed in a future work, in which even $\{ z_i(L_i) \}_i$ is a derived, not free, set of parameters.

\subsection{Comparing the model with cosmological observations}\label{observations}

In this section, we aim to find out how the model with $\alpha(z)$ given by Eq.\ (\ref{DensityAlongOurLineOfSight}) fits the main cosmological data sets: the cosmic microwave background anisotropy, the position of the baryon acoustic oscillation peak inferred from the galaxy distribution, the magnitude-redshift relations of type Ia supernovae, the local measurements of the Hubble constant and the Big Bang nucleosynthesis. For a thorough and open-minded review of the current observational status, see Ref.\ \cite{Sarkar:2007cx}.

A good baseline for the studies is the model by Blanchard, Douspis, Rowan-Robinson and Sarkar \cite{Blanchard:2003du,Blanchard:2005ev}, and especially its refined version by Hunt and Sarkar \cite{Hunt:2004vt,Hunt:2007dn,Hunt:2008wp}. This is an FRW model with the following basic properties: no dark energy or $\Lambda = 0$, a mixture of cold and hot dark matter $\Omega_{{\rm CDM}}=0.8$ \& $\Omega_{{\rm HDM}}=0.1$, flat spatial geometry $k=0$, a low Hubble constant $h = 0.44$ and a bump in the primordial perturbation spectrum at the scales $k \sim 0.01 ... 0.1 ~ h {\rm Mpc}^{-1}$, which could arise e.g.\ from phase transitions during inflation \cite{Adams:1997de}; we shall call this the \emph{linear} CHDM model. The model fits the CMB angular power spectra better than the standard $\Lambda$CDM model. It is also consistent with the nucleosynthesis constraints and with most of the features in the matter power spectrum of the galaxy distribution surveys. However, the model seems to fail in matching the observed magnitude-redshift relations of type Ia supernovae and the detected position of the baryon acoustic oscillation peak in the matter power spectrum and it is also in disagreement with the local measurements of the Hubble constant. For a more thorough description of the linear CHDM model, see \cite{Hunt:2007dn}.

A candidate for hot dark matter would be three species of massive neutrinos with $\sum_i m_{\nu_i} \sim 1.8 ~ \rm{eV}$, yielding $\Omega_{{\rm HDM}} = \Omega_\nu \simeq 0.1$. This is well below the present experimental upper bound of $m_{\nu} = 2.3 ~ \rm{eV}$ \cite{Yao}, but is expected to be detectable in the forthcoming KATRIN $\beta$-decay experiment \cite{Drexlin:2005zt}. Note that the sometimes quoted tighter upper bounds from cosmological observations depend heavily on the priors \cite{Elgaroy:2003yh}. As neutrinos with mass higher than $m_\nu \gtrsim 0.3 ~ \rm{eV}$ have been shown to cluster appreciably \cite{Singh:2002de}, their existence should not alter the assumption that the detectable light travels in almost empty space today.

We consider a universe that is similar to the linear CHDM model at large redshifts, but instead of describing its optical properties with the standard FRW relations, we use the generalized Dyer-Roeder equations (\ref{dAmodifiedDR})--(\ref{dLmodifiedDR}) to determine the observable distance-redshift relations. Accordingly, the growth of small deviations from homogeneity is assumed to follow linear perturbation theory, so apart from the changes in the associated distance measures, the perturbation spectra should be more or less unaltered. We call this the \emph{nonlinear} CHDM model. We pose five observational tests for this model:
\begin{enumerate}
\item To obtain the same value for the angular diameter distance to the last scattering surface as in the linear CHDM model.
\item To match the position of the baryon oscillation peak in the matter power spectrum.
\item To fit the observed magnitude-redshift relations of the type Ia supernovae.
\item To explain the locally measured value of the Hubble constant.
\item To remain within the limits on the baryon density set by the BB nucleosynthesis.
\end{enumerate}
By passing these tests successfully, the nonlinear CHDM model would explain the main cosmological data sets that support accelerated expansion. In Sects.\ \ref{CMB}--\ref{nucleosynthesis}, we discuss how the model survives these tests and the results are combined in Sect.\ \ref{combinedresults}.

\subsubsection{The cosmic microwave background}\label{CMB}

For the values $\bar{\Omega}_B=0.1$, $\bar{h}=0.44$ and $\bar{\Omega}_0=1$, the fitting formula of Hu and Sugiyama \cite{Hu:1995en} yields $z_{{\rm dec}} = 1101.35$ as the decoupling redshift, for which we use the approximation $z_{{\rm dec}} \simeq 1100$ throughout.

The fit of the linear CHDM model to the WMAP $3$-year data has been thoroughly performed by Hunt and Sarkar \cite{Hunt:2007dn} so we only need to consider how the nonlinearities modify this. As the nonlinear structures appear only in late times, it is essentially only the comoving angular diameter distance $d_c(z) \equiv (1+z) d_A(z)$ and the late integrated Sachs-Wolfe effects that can change. However, as the model already fits the CMB data \emph{without} the nonlinear structures, it is desirable that adding the nonlinearities would not alter the distance to the last scattering surface, $d_c(1100)$. Physically, this is possible because the negative spatial curvature and the faster Hubble expansion in the voids have the opposite effects on the comoving angular diameter distance $d_c(z)$, so with a suitable value for the new parameter $z_0$, their effects can cancel.

By determining $d_c(z)$ numerically from Eq.\ (\ref{dAmodifiedDR}), and $\bar{d}_c(z)=\bar{d}_L(z)/(1+z)$ from Eq.\ (\ref{lumiFRWgeneral}), we find that the value for $z_0$ that sets $d_c(1100)$ and $\bar{d}_c(1100)$ identical, is $z_0=0.347$. Values in the range $z_0 = 0.30...0.39$ give less than 1\% discrepancy between $d_c(1100)$ and $\bar{d}_c(1100)$. Although equal around $z \approx 1100$, the functions $d_c(z)$ and $\bar{d}_c(z)$ are not equal at lower redshifts.

Varying the Hubble constant does not change the limits on $z_0$ as $\bar{H}_0$ has essentially the same effect on both the "background"\ value $\bar{d}_c(1100)$ and the value $d_c(1100)$ which takes into account the nonlinearities. To conserve the successful angular power spectrum found in Ref.\ \cite{Hunt:2007dn}, we fix the FRW Hubble constant to the value $\bar{H}_0 = 44~{\rm{kms^{-1}Mpc^{-1}}}$.

Within the linear perturbation theory, a flat matter dominated FRW model does not have a late integrated Sachs-Wolfe (LISW) effect, because the perturbed gravitational potentials do not evolve in time, $\dot{\Phi}=0$. However, this does not necessarily hold in a universe with nonlinear structures if the CMB light has propagated mostly in voids. Indeed, with $\dot{\Phi}$ potentially non-zero along our line of sight, one would expect a non-trivial LISW effect from nonlinear inhomogeneities also in a matter dominated universe with initially flat FRW geometry. On the other hand, as the LISW affects the angular power spectrum only at low multipoles $l$, its contribution to the goodness of the overall fit could be insignificant due to the large cosmic variance $\propto (2l+1)^{-1}$. Moreover, it might in fact be that the effects of the local structures dominate the angular power at the lowest multipoles \cite{Rakic:2007ve,Gurzadyan:2007ic}. It is also possible that, even if the selection effects were significant in the supernova observations, they may be less important in the CMB observations. Altogether, along with other more involved issues, we postpone a quantitative discussion of the LISW to a future work.

\begin{figure}[tbh]
\begin{flushleft}
\includegraphics[angle=-90,scale=0.6]{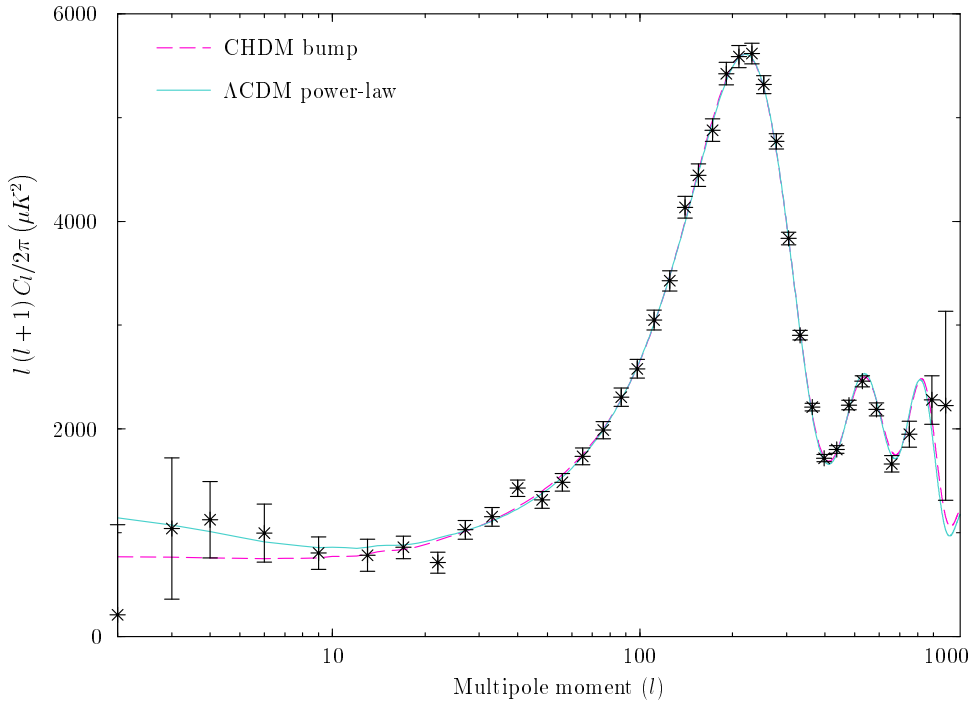}
\caption{\label{Powerspectrum} Binned WMAP data points in the temperature power spectrum with theoretical predictions from the best fit CHDM and $\Lambda$CDM models; figure from Hunt and Sarkar \cite{Hunt:2007dn}.}
\end{flushleft}
\vspace{-0.6cm}
\end{figure}

Apart from the potential modifications due to the LISW effect at small $l$, the angular power spectrum of the nonlinear CHDM model with $z_0=0.347$ is plotted in Fig.\ \ref{Powerspectrum}.

\subsubsection{Galaxy distribution}\label{galaxysurveys}

As long as the nonlinear inhomogeneities only affect light, the evolution of the linear perturbations remains the same as in the linear CHDM model. In the case of the CMB power spectrum this seems obvious, since the relevant evolution of the perturbations happened at $z<1100$ when the nonlinearities were negligible. However, as the galaxy distribution also reflects the dynamics of the late universe, it is not clear that the nonlinear inhomogeneities would not play a dynamical role here. We nevertheless neglect the potential effects of the nonlinearities on the evolution of the perturbations, but remind that, in a more realistic treatment, these effects should be taken into account in the matter power spectrum.

As shown in Refs.\ \cite{Blanchard:2005ev,Hunt:2007dn}, the linear CHDM model fits the matter power spectrum apart from the position of the baryon acoustic oscillation peak. As in the CMB angular power spectrum, the main modification caused by the nonlinearities is the change in the distance-redshift relations. This opens up a possibility that the nonlinear CHDM model would fit the position of the baryon acoustic oscillation peak. To study this, we use the data from the Sloan Digital Sky Survey provided in Ref.\ \cite{Eisenstein:2005su}.

The detected position of the baryon acoustic oscillation peak in the matter power spectrum constrains the distance to the redshift $z=0.35$, corresponding to the acoustic oscillation scale today. More specifically, due to the different scaling of radial and angular distances, the exact distance measure constrained by the data reads as \cite{Eisenstein:2005su}:
\begin{equation}\label{DeeVeeDef2}
D_V(z) \equiv \left[ \frac{z d_c^2(z)}{H(z)} \right]^{1/3}~.
\end{equation}
As the actual observations are made in the redshift space and the distance-redshift relation depends on the cosmological model, the "measured"\ values quoted in \cite{Eisenstein:2005su} are prior-dependent. We take this into account by using the more conservative $2\sigma$ error limits. The value of the distance (\ref{DeeVeeDef2}) to the redshift $z=0.35$ deduced from the SDSS data is
\begin{equation}\label{DeeVee2sigma}
D_V(0.35) = 1370 \pm 150 ~ {\rm Mpc} ~,
\end{equation}
where the limits corresponding to $2\sigma$ errors have been projected from Fig.\ 7 of Ref.\ \cite{Eisenstein:2005su}. With $H_0 = 66~{\rm{kms^{-1}Mpc^{-1}}}$ fixed, we find that the constraint set by Eq.\ (\ref{DeeVee2sigma}) on the transition redshift parameter in Eq.\ (\ref{DensityAlongOurLineOfSight}) is $z_0 > 0.25$, whereas the best fit value from the CMB analysis of Sect.\ \ref{CMB}, $z_0=0.347$, gives $D_V(0.35) = 1480 ~ {\rm Mpc}$. Fixing $z_0=0.347$ requires $H_0 = 71~{\rm{kms^{-1}Mpc^{-1}}}$ or $\bar{H}_0 = 47~{\rm{kms^{-1}Mpc^{-1}}}$ to give the best fit value $D_V(0.35) = 1370 ~ {\rm Mpc}$.

In the case that the statistical properties of the primordial perturbations are the same at every spatial location in the universe, the baryon oscillation peak in the matter spectrum and the first peak of the CMB angular power spectrum should correspond to the same physical feature just seen at different spatial locations and at different epochs. The physical scale associated with these oscillations is set by the sound horizon at recombination. As the distance (\ref{DeeVee2sigma}) measures how large this length scale appears in the galaxy distribution, it is possible to combine it with the information of the associated scale of the peak in the CMB spectrum to make an additional observational test. The quantitative measure employed in Ref.\ \cite{Eisenstein:2005su} for this is defined as:
\begin{equation}\label{R035def}
R_{0.35} \equiv \frac{D_V(0.35)}{d_c(1100)} ~.
\end{equation}

It should be again noted that due to the prior-dependency of the distance-redshift relations, Eisenstein et.\ al.\ quote several "measured"\ values for $R_{0.35}$. As $d_c(1100) = 13190 ~ {\rm Mpc}$ is the best fit value to the CMB data in the CHDM model, Eq.\ (\ref{DeeVee2sigma}) yields:
\begin{equation}\label{R0352sigma}
R_{0.35} = 0.104 \pm 0.011~,
\end{equation}
where the limits should roughly correspond to $2\sigma$ errors in the CHDM model. We find that the values of $z_0$ corresponding to these limits are: $z_0 = 0.29...0.83$, with $z_0=0.53$ yielding the best fit value $R_{0.35} = 0.104$. As $R_{0.35}$ is a ratio of distances, it is independent of $H_0$.

Albeit the nonlinear CHDM model with $\alpha(z)$ given by Eq.\ (\ref{DensityAlongOurLineOfSight}) is not as concordant with the SDSS data as with the other data sets, the value $z_0=0.35$ still falls within the experimental limits. Moreover, the results represent a huge improvement compared to the linear CHDM model, which yields the values $\bar{D}_V(0.35)=1760 ~ {\rm Mpc}$ and $\bar{R}_{0.35}=0.134$ that are both way off the rather conservative limits of Eqs.\ (\ref{DeeVee2sigma}) and (\ref{R0352sigma}).

\subsubsection{Type Ia supernovae}\label{supernovae}

We use the Riess et.\ al.\ gold sample of $182$ type Ia supernovae \cite{Riess:2006fw} to study how the nonlinear CHDM model fits the SN observations. For this, we use the conventional $\chi^2$ test:
\begin{equation}\label{chisquared}
\chi^{2} \equiv \sum_{n=1}^{182} \left( \frac{m^{{\rm{obs}}}(z_n)-m(z_n)}{\sigma_n} \right)^2~,
\end{equation}
where $m(z_n) \equiv 5 \log_{10}(d_L(z_n) /{\rm Mpc})+ 25$ is the theoretical prediction for the magnitude and $\sigma_n$ is the estimated error for the measured magnitude $m^{{\rm{obs}}}(z_n)$ of a source at the redshift $z_n$.

Just like the flat $\Lambda$CDM model with $\bar{\Omega}_\Lambda$ and $\bar{H}_0$ as free parameters, the nonlinear CHDM model has two free parameters: the transition redshift $z_0$ and, say, the observable Hubble constant $H_0$. There is, however, a perfect degeneracy between $H_0$ and the maximum intrinsic luminosity of the type Ia supernovae, $L_{{\rm Ia}}$, so the uncertainty in the value of $L_{{\rm Ia}}$ reflects a similar uncertainty in the value of $H_0$ constrained from this data. For this reason, the combination ($H_0$,$L_{{\rm Ia}}$) is sometimes marginalized in the data analysis but we choose to instead use the fixed value for $L_{{\rm Ia}}$ employed by Riess et.\ al. in Ref.\ \cite{Riess:2006fw}.

By letting both parameters vary, we find the best fit values $H_0 = 65.5~{\rm{kms^{-1}Mpc^{-1}}}$ and $z_0=0.39$ that yield $\chi^2 = 161.1$. A little inspection in the parameter space shows that there is degeneracy between $z_0$ and $H_0$, such that one can obtain good two-parametric fits ($\Delta \chi^2 \lesssim 5$) roughly in the range: $z_0 = 0.2...0.8$. Keeping instead $H_0$ fixed to the value $66~{\rm{kms^{-1}Mpc^{-1}}}$, we find that the best fit transition redshift is $z_0=0.34$, yielding $\chi^2 = 161.4$. The increase $\Delta \chi^2 =2.7$ corresponds to $2\sigma$ errors in a one-parametric fit and yields for the limits: $z_0=0.34 \pm^{0.10}_{0.08}$. The absolute fit $161.4/182 = 0.89$ is good and represents a tremendous improvement to the linear CHDM model where $\chi^2 = 1098$ for $\bar{h}=0.44$ and even the best fit value $\bar{h}=0.54$ gives $\chi^2=283.4$.

\begin{figure}[tbh]
\begin{center}
\includegraphics[width=15.0cm]{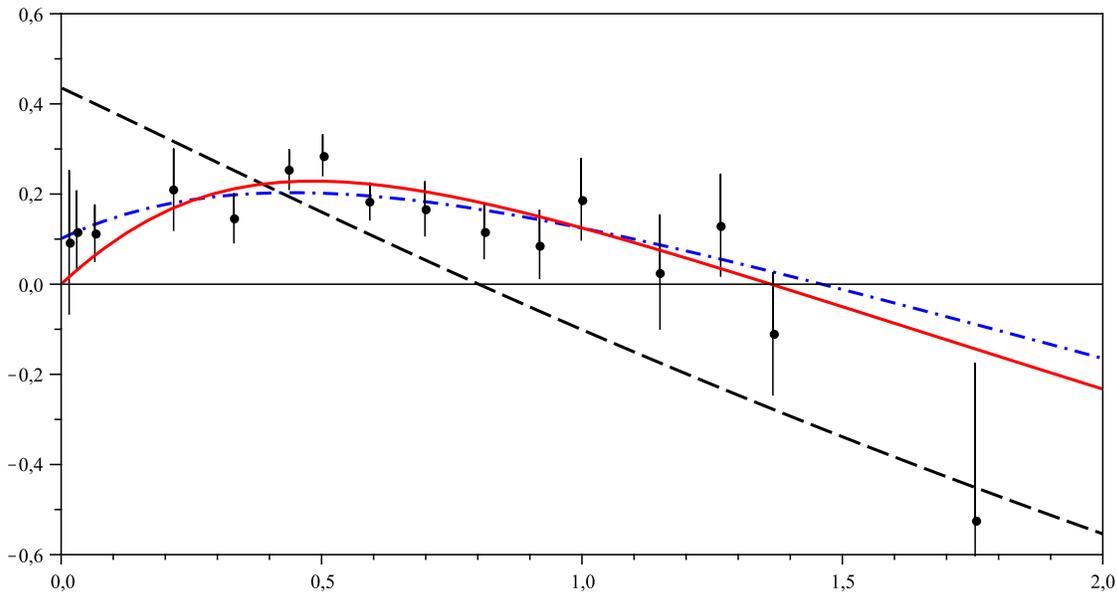}
\caption{16 binned data points from the Riess et.\ al.\ gold sample in the residual magnitude-redshift diagram with theoretical predictions from three different models: the linear CHDM model with $\bar{h}=0.54$ (dashed black curve), the $\Lambda$CDM model with $\bar{\Omega}_\Lambda=0.67$ and $\bar{h}=0.63$ (dashed dotted blue curve), and the nonlinear CHDM model with $z_0=0.34$ and $h=0.66$ (solid red curve). We have subtracted the $\bar{h}=0.66$ freely expanding Milne universe relation from the curves to produce the residual diagram, so only the $h=0.66$ nonlinear CHDM model intersects the origin.}\label{figuuri}
\end{center}
\vspace{-0.6cm}
\end{figure}

For comparison, the best fit parameters of the $\Lambda$CDM model $\bar{\Omega}_\Lambda=0.67$ and $\bar{h}=0.63$ yield $\chi^2 = 158.8$. However, for the concordance values $\bar{\Omega}_\Lambda=0.73$ and $\bar{h}=0.7$ the fit is much worse: $\chi^2 = 305.2$, but this can be fixed by changing the assumed value of $L_{{\rm Ia}}$ or equivalently changing $\bar{h}$. Indeed, we find that the value $\bar{h}=0.642$ gives the best fit for the concordance $\bar{\Omega}_\Lambda=0.73$ $\Lambda$CDM model with $\chi^2 = 162.5$. Overall, the conclusion from comparing the results between the nonlinear CHDM model and the $\Lambda$CDM model is that the goodness of their fits to the Riess et.\ al.\ data are essentially equal. In Fig.\ \ref{figuuri}, we plot the residual magnitude-redshift relations for three different models: the linear CHDM model with $\bar{h}=0.54$; the $\Lambda$CDM model with $\bar{\Omega}_\Lambda=0.67$ and $\bar{h}=0.63$; and the nonlinear CHDM model with $z_0=0.34$ and $h=0.66$.

\subsubsection{The Hubble Key Project}\label{HubbleKey}

The best fit values obtained from the different data sets of Sects.\ \ref{CMB}, \ref{galaxysurveys} and \ref{supernovae} for the observable Hubble constant are all consistent with the low redshift measurements of the expansion rate that should specifically probe the observable Hubble constant; the value $H_0 = 66~{\rm{kms^{-1}Mpc^{-1}}}$ sits firmly within the limits of the final results from the Hubble Key Project: $H_0 = 72 \pm 8 ~ {\rm{kms^{-1}Mpc^{-1}}}$ \cite{Freedman:2000cf} or $H_0 = 62.3 \pm 5.2 ~ {\rm{kms^{-1}Mpc^{-1}}}$ \cite{Sandage:2006cv}. In addition, there are persistent indications that one gets systematically lower values for the Hubble constant by observing objects at higher redshifts \cite{Sarkar:2007cx}. For the standard FRW cosmology this feature is an anomaly, while in the nonlinear CHDM model it is a prediction: towards higher redshifts, the optically perceived expansion should be closer to the (lower) FRW expansion rate due to the fact that the earlier universe was more homogeneous.

\subsubsection{Age of the high redshift quasar APM $08279 + 5255$}\label{quasars}

The particularly well-observed high redshift quasar APM $08279 + 5255$ \cite{Hasinger:2002wg,Komossa:2002cn} at $z=3.91$ with estimated age $2...3~{\rm Gyr}$ has been proposed as a litmus test for every cosmological model \cite{Jain:2005gu}. For several cosmological models, most notably the homogeneous standard $\Lambda$CDM model where $t_{{\rm{age}}}(3.91)=1.7~{\rm Gyr}$ and almost all the dark energy models, the observed quasar implies an age problem: the universe appears to be younger than its contents.

The considered nonlinear model offers a resolution to the age problem: if the observed light from the quasar propagates in voids that expand faster than the average, the light redshifts more rapidly than expected from the stretching of the global FRW scale factor, implying that the quasar is seen in an older universe than predicted by the age-redshift relation of the homogeneous FRW model.
An order of magnitude estimate can be obtained by approximating that the space along our line of sight to the quasar has expanded freely. This increases the age at a given redshift $z$ relative to the FRW value by a factor $\sqrt{1+z}$, yielding $t_{{\rm{age}}}=3~{\rm Gyr}$ for the age of the universe at $z=3.91$, consistent with the estimated age $2...3~{\rm Gyr}$ of the quasar APM $08279 + 5255$.

\subsubsection{The Big Bang Nucleosynthesis}\label{nucleosynthesis}

Finally, we comment on the nucleosynthesis. The overall result from deducing the abundances of the light elements from the spectra of stars and galaxies is that the baryon density falls within the range $0.017 \leq \bar{\Omega}_B \bar{h}^2 \leq 0.024$ at $95$\% confidence level \cite{Fields:2006ga}. As long as the early universe was nearly homogeneous and the constants of nature do not vary, these limits are model-independent. Therefore, the observed value should not be affected by the nonlinear inhomogeneities in the late-time universe so that the value for the baryon density, $\bar{\Omega}_B \bar{h}^2 = 0.1 \cdot 0.44^2 = 0.0194$, falls within the above quoted observational limits also in the nonlinear CHDM model.

\subsubsection{A concordant nonlinear CHDM model with no dark energy}\label{combinedresults}

By combining the results from the analysis of the different data sets in Sects.\ \ref{CMB}--\ref{nucleosynthesis}, we obtain a concordant cosmological model with the following properties: the observable Hubble constant $h = 0.66$, the transition redshift to the era when nonlinear structures become dominant $z_0=0.35$, baryon proportion $\bar{\Omega}_B=0.1$, cold dark matter proportion $\bar{\Omega}_{{\rm CDM}} = 0.8$, hot dark matter proportion $\bar{\Omega}_{{\rm HDM}} =0.1$ and no dark energy $\Lambda=0$. All the data sets are consistent with these values and, apart from the SDSS data, the concordance value for $z_0$ is actually very close to the best fit value of each separate data set; see Table 1. Moreover, the concordance value for $H_0$ yields $t_0 = H_0^{-1} = 14.8~{\rm Gyr}$ as the age of the universe, consistent with the lower bound determined from the astronomical estimates for the age of the oldest globular clusters: $11.2~ {\rm Gyr}$ (at $95\%$ confidence level) \cite{Krauss}. The main properties of the best fit nonlinear CHDM model are summarized in Table 2.

\vspace{0.4cm}
\begin{table}[tbh]
\begin{center}
\begin{tabular}{lcccc}
\hline  Description     & CMB            & BAO peak           &  Type Ia SNe       & Concordance value    \\ \hline
 Transition redshift $z_0$           & $0.35\pm^{0.04}_{0.05}$  & $0.53 \pm_{0.24}^{0.30}$ & $0.34 \pm^{0.10}_{0.08}$ &  $0.35$              \\
\hline
\end{tabular}
\caption{The constraints on the transition redshift from each data set and the concordance value.}
\end{center}
\end{table}
\vspace{-0.6cm}
\begin{table}[tbh]
\begin{center}
\begin{tabular}{lcc}
\hline Description & Parameter  &  Concordance value     \\ \hline
Transition redshift & $z_0$ & $0.35$           \\
Observable Hubble constant      & $h$        & $0.66$           \\
FRW Hubble constant    & $\bar{h}$& $0.44$           \\
Age of the universe &  $t_0$ & $14.8~{\rm Gyr}$ \\
Baryon proportion      & $\bar{\Omega}_B$          & $10\%$           \\
Cold dark matter proportion     & $\bar{\Omega}_{{\rm CDM}}$& $80\%$            \\
Hot dark matter proportion  & $\bar{\Omega}_{{\rm HDM}}$& $10\%$           \\
\hline
\end{tabular}
\end{center}
\begin{center}
\caption{Best fit parameters for the nonlinear CHDM model with no dark energy. Only one of the three parameters $h$, $\bar{h}$ and $t_0$ is independent. It is noteworthy that a phenomenologically similar model was found by Wiltshire et.\ al.\ \cite{Leith:2007ay}, though with different interpretation for how the voids give rise to the acceleration \cite{Wiltshire:2007fg,Wiltshire:2007jk,Leith:2007ay,Wiltshire:2007zh,Wiltshire:2007zj,Wiltshire:2008sg}.}
\end{center}
\end{table}

\subsection{Discussion of the fits}\label{results}

For consistency, we have tested various forms for the structure formation function $\alpha(z)$, in addition to the form (\ref{DensityAlongOurLineOfSight}) employed in the data analysis in Sect.\ \ref{observations}. The generic outcome seems to be that the exact form of the function $\alpha(z)$ is not crucial to fit the data as long as it changes roughly from $\alpha(1100) \approx 1$ to $\alpha(0) \approx 0$. Neither does the fit require that the observed light should travel in totally empty space today: it is possible to have nonzero present-day average matter density on our line of sight $\alpha(0) \simeq \mathcal{O} (0.1)$ and still fit the data sets, although with slightly different best fit values for $z_0$, $t_0$ and the new parameter $\alpha(0)$. An important implication is that the seemingly worrisome feature of the matter density along our line of sight becoming exactly zero today ($\Longleftrightarrow \alpha(0)=0$) is not necessary, but we have chosen the one-parametric form (\ref{DensityAlongOurLineOfSight}) for simplicity.

As an example, the more sophisticated two-parametric form
\begin{equation}\label{hooztanh}
\alpha(z)=  \left( \frac{1 + \tanh ( z/z_0 - \phi )}{2}\right)^2 \left(\frac{5 - \tanh ( z/z_0 - \phi )}{4}\right)^2 ~,
\end{equation}
with the values $z_0=0.35$ and $\phi=1.46$ yields an even more concordant fit to the combined data sets; it gives $D_V(0.35)=1370 ~{\rm Mpc}$ for (\ref{DeeVee2sigma}), $R_{0.35}=0.104$ for (\ref{R0352sigma}), $\chi^2=163$ for the goodness of the supernova fit and $d_c(1100)/\bar{d}_c(1100)=1.000$ or no deviation from the distance to the LSS in the best fit linear CHDM model.

Due to the approximations employed in Sects.\ \ref{dyerroeder} and \ref{derivation} to enable an analytic treatment, we expect that the numerical values of the best fit parameters can represent the effect of the inhomogeneities only at a relatively crude level. In a more realistic treatment, the analytic approximations should be replaced with more thorough numerical procedures and perform the fits to all the data sets from scratch -- e.g.\ by making the necessary data calibrations using a nonlinear model as the baseline instead of the linear $\Lambda$CDM model. This concerns especially the matter power spectrum, for which the full analysis requires considerably more effort and is therefore postponed to a future work. Instead, due to the mutual degeneracy between the nonlinear structure formation and the cosmological constant $\Omega_\Lambda$, evident e.g.\ in Eqs.\ (\ref{lumiFRW}) and (\ref{lumiDRmodified}), the fit does not gain a noteworthy improvement by keeping $\Omega_\Lambda$ in the analysis.

An inevitable consequence of considering observables that represent averages over different directions on the sky is the mathematical resemblance between the model of Sect.\ \ref{themodel} and a geocentric LTB bubble. Naturally, with photons from different directions propagating through different void profiles, the isotropic Dyer-Roeder distance-redshift relations can only represent idealized ensemble averages. The actual observations should thus contain angular variations that are relatively smaller when making observations over distances much greater than a homogeneity scale and hence the largest at low redshifts. Indeed, considerable amount of scatter with unidentified cause has been observed e.g.\ in the magnitude-redshift relations of the type Ia supernovae and importantly, especially at low redshifts \cite{Wang:2006cq,Schwarz:2007wf,Seikel:2007pk}; for detailed analysis of scatter in HKP data, see \cite{McClure:2007vv}. The observed anomalies at the low multipoles of the angular power spectrum represent corresponding features in the CMB data \cite{Rakic:2007ve,Gurzadyan:2007ic}. Moreover, Gurzadyan et.\ al.\ have identified features in the randomness of the raw CMB data that could originate from voids \cite{Gurzadyan:2008va}. Altogether, it appears a promising option that these apparent peculiarities would in fact be caused, at least partially, by the nonlinear inhomogeneities and thus provide evidence for their cosmological role addressed in this work. However, due to our lack of knowledge from the detailed (dark) matter distribution and expansion rate, it seems difficult to make precise quantitative predictions for the angular deviations. An important task is to search for correlations between the apparent anomalies of the different data sets.

\section{Conclusions}\label{konkluusiot}

We have studied the effect of structure formation on the observable distance-redshift relations by generalizing the Dyer-Roeder method to allow for two important physical properties of the universe: inhomogeneities in the expansion rate and the growth of nonlinear structures. We noted that light propagation through voids can cause the expansion along our line of sight to increase from flat dust FRW-expansion at high redshifts to free expansion at low redshifts and thereby have a similar effect on the observed distance-redshift relations as dark energy. Conjecturing such a transition, we showed that the inhomogeneities offer a natural physical explanation for the cosmological observations without dark energy even with a single physically well-grounded phenomenological parameter, the transition redshift $z_0$ to the era when the matter density becomes negligible along our line of sight or when the detectable light starts to propagate in freely expanding space.

Indeed, applying the generalized Dyer-Roeder method in Sect.\ \ref{observations}, we found a simple inhomogeneous model that fits the observations from the CMB anisotropy, the position of the baryon acoustic oscillation peak inferred from the SDSS data, the magnitude-redshift relations of type Ia supernovae, the local Hubble flow and the Big Bang nucleosynthesis, plus is potentially consistent with the estimated age of the quasar APM $08279 + 5255$ at $z=3.91$. The main features of the resulting concordant model are: 90\% dark matter, 10\% baryons, $z_0=0.35$ as the transition redshift to the void-dominated era, the values $h=0.66$ and $\bar{h}=0.44$ for the observable and FRW Hubble constants that yield $t_0 = 14.8~{\rm Gyr}$ as the age of the universe, consistent with astronomical estimates, and no dark energy of any kind or $\Lambda=0$. At large redshifts, this nonlinear CHDM model is similar to the linear CHDM model by Hunt and Sarkar \cite{Hunt:2004vt,Hunt:2007dn}. A non-trivial point in the best fit model is that the distance-redshift relation is not a reproduction of the corresponding relation in the $\Lambda$CDM model. The model is based on inhomogeneities of the observed kind so, unlike a large void in the local matter density, there is no violation of the cosmological principle: it is our special location in time, not in space, that is required for the acceleration. Therefore, a prediction is that \emph{every} observer in the present-day universe should perceive the inhomogeneity induced acceleration.

Although homogeneous models can also provide a phenomenological fit to the cosmological data by assuming dark energy, they entail various issues that can be formulated into five questions:
\begin{enumerate}
\item What justifies the use of the FRW metric to model light propagation in the real universe?
\item Why would dark energy appear at a very low temperature $T \simeq 4 ~ {\rm K}$ if it has had the huge temperature range from $T_{{\rm P}} \simeq 10^{32}~ {\rm K}$ to $T_0 \simeq 2.7$ K available? An expectation for the probability of a quantum field to arise at a temperature $T$ is $P \sim e^{-T_{{\rm P}}/T}$.
\item Why would the effects of dark energy appear when large voids start to form?
\item Why would $\Lambda$ be fine-tuned such that the universe undergoes nearly free expansion today, $H_0 t_0 =\frac{ 2}{3 \sqrt{\Omega_\Lambda}} {\rm {arsinh} } \sqrt{\frac{\Omega_\Lambda}{1-\Omega_\Lambda}} \simeq 1$, whereas a slightly bigger $\Lambda$ would yield $H_0 t_0 >1$? This issue was pointed out by Kutschera and Dyrda in Ref.\ \cite{Kutschera:2006bh}.
\item If nonzero, why would the cosmological constant be so tiny, $\Lambda \sim 10^{-123}~G^{-1}$, whereas naive dimensional analysis suggests $\Lambda \sim G^{-1}$?
\end{enumerate}
The structure formation offers a smart solution to the coincidence problem: if induced by the voids, the onset of the perceived acceleration naturally coincides with the formation of the voids. Furthermore, the observed $H_0 t_0 \sim 1$ corresponds to the characteristic expansion rate of freely expanding voids, $t \sim t_0/2$ is a natural time scale for gravity to form structures on the relevant length scales and $\Lambda = 0$ is compatible with the data if void formation is responsible for the acceleration. In addition, there are various observed features in the universe that are anomalies for the standard FRW cosmology, such as the cold spot in the CMB, the low multipoles of the CMB, the CMB ellipticity and the scatter in the supernova data, but could be explained naturally by the anisotropy of the cosmic void distribution in the sky; a more exhaustive list of such features can be found in Ref.\ \cite{Wiltshire:2007zj}. As we have only discussed how to construct models of the observed inhomogeneous universe within relativistic classical gravity, the answer to the long-standing debate of \emph{why} would $\Lambda=0$ is naturally beyond the scope here.

The transition to the nonlinear era, roughly given by $z_0=0.35$, might seem to happen rather late, only $\Delta t = 5.4 ~{\rm Gyr}$ back in time, whereas the oldest stars have formed already $\Delta t > 12 ~{\rm Gyr}$ ago. However, there are in fact several natural reasons to expect such a seemingly late transition. Firstly, while traveling through structures that are still collapsing in earlier times, light is blueshifted which can partially counterbalance the excess redshift caused by the light propagation in voids. Later, however, when the structures have virialized also at larger scales, there are more and more structures which neither collapse nor expand. Those photons that are able to penetrate these virialized overdense regions, then do not blueshift anymore, diminishing this partial counterbalance at late times. Secondly, the large amount of weakly interacting dark matter in the universe prefers to form large halos which cluster into smaller clumps later than the luminous matter which nonetheless indicates the clumping to us. It also appears likely that the voids can grow large enough only in the late universe. Finally, when contrasting the acceleration from structure formation with dark energy driven acceleration, it is essential to note that the rate at which gravitational phenomena happen is largely independent of the temperature, whereas non-gravitational interactions happen faster at higher temperatures. Therefore, while it seems extremely difficult to explain why a quantum field would arise at $T \simeq 4~{\rm K}$ especially if it has had the huge temperature range from $T_{{\rm{P}}} \simeq 10^{32}~{\rm K}$ to $T_0 \simeq 2.7~{\rm K}$ available, it is natural that the inhomogeneity induced acceleration would start as "late"\ as $\Delta t \simeq 7 ~{\rm Gyr}$ ago (corresponding to redshift $z \simeq 0.52$ that can be read off e.g.\ from Fig.\ \ref{figuuri}).

The analytic approach employed in this work necessitates approximations, so the precision of the numerical values for the best-fit parameters presented here is not expected to be as high as an ideal analysis could yield. Thus, it is difficult to say anything conclusive about the subtler features of the concordant nonlinear CHDM model, such as the bump in the primordial spectrum or the small hot dark matter component. Instead, given the physically well-grounded foundations of the proposed conjecture about light propagating mostly in voids, the observational evidence for the nonlinear structures as well as the large volume taken up by voids, the concordant fit to the cosmological data and the offered smart solutions to some of the perhaps most puzzling theoretical problems in the modern cosmology, it seems perfectly conceivable that the main physical implications -- most notably that dark energy may not exist -- could be correct.

An important future test is to study quantitatively the observed angular deviations in various data that are an anomaly for the homogeneous standard $\Lambda$CDM model, but a predicted feature in the inhomogeneous models arising from the anisotropic void distribution. Another crucial test is to demonstrate the proposed conjecture within less approximative solutions of general relativity. Ideally, a thorough derivation of the conjecture from theory would involve a numerical computation employing the full Einstein equations starting with some initial perturbations in the early universe. More realistically, the feasible next step is to replace some of the analytic approximations with more careful numerical procedures; we leave this analysis to a future work, where e.g.\ the transition redshift $z_0$ will be a derived, not phenomenological, parameter. Although potentially laborious, these tests should be realizable with present-day knowledge, so the proposed conjecture is certainly falsifiable.

\acknowledgments{This paper would not exist without the altruistic help of Maria Mattsson, whom I heartily thank. I also thank Subir Sarkar, Kimmo Kainulainen and Syksy R\"as\"anen for valuable discussions and Tomi Koivisto, Kari Enqvist, Morad Amarzguioui, H{\aa}vard Alnes, David Wiltshire and Dominik Schwarz for helpful comments. The author is financially supported by the Magnus Ehrnrooth Foundation. This work was also supported by the European Union through the Marie Curie Research and Training Network ``UniverseNet'' (MRTN-CT-2006-035863).}


\begin{thebibliography}{99}

\bibitem{Zeldovich}
  Ya.~B.~Zel'dovich,
  ``Observations in a Universe Homogeneous in the Mean'',
  Soviet Ast.\ (1964) 8, 13

\bibitem{Bertotti}
  B.~Bertotti,
  ``The Luminosity of Distant Galaxies'',
  Proc.\ R.\ Soc.\ London A {\bf 294} (1966) 195.

\bibitem{Gunn1}
 J.~E.~Gunn
 ``On the Propagation of Light in Inhomogeneous Cosmologies. I. Mean Effects'',
 Astrophys.\ J.\ {\bf 150} (1967) 737

\bibitem{Kantowski1}
 R.~Kantowski
 ``Corrections in the Luminosity-Redshift Relations of the Homogeneous Friedmann Models'',
 Astrophys.\ J.\ {\bf 155} (1969) 89

\bibitem{DyerRoeder1}
 C.~C.~Dyer and R.~C.~Roeder
 ``The Distance-Redshift Relation for Universes with no Intergalactic Medium'',
 Astrophys.\ J.\ {\bf 174} (1972) L115

\bibitem{DyerRoeder2}
 C.~C.~Dyer and R.~C.~Roeder
 ``Distance-Redshift Relations for Universes with Some Intergalactic Medium'',
 Astrophys.\ J.\ {\bf 180} (1973) L31

\bibitem{Hoyle:2003hc}
  F.~Hoyle and M.~S.~Vogeley,
  ``Voids in the 2dF Galaxy Redshift Survey'',
  Astrophys.\ J.\  {\bf 607} (2004) 751
  [arXiv:astro-ph/0312533].

\bibitem{Gott:2003pf}
  J.~R.~I.~Gott {\it et al.},
  ``A Map of the Universe'',
  Astrophys.\ J.\  {\bf 624} (2005) 463
  [arXiv:astro-ph/0310571].

\bibitem{Tikhonov:2007di}
  A.~V.~Tikhonov,
  ``Voids in the SDSS Galaxy Survey'',
  Astron.\ Lett.\  {\bf 33} (2007) 499
  [arXiv:0707.4283 [astro-ph]].

\bibitem{vonBendaBeckmann:2007wt}
  A.~M.~von Benda-Beckmann and V.~Mueller,
  ``Void Statistics and Void Galaxies in the 2dFGRS'',
  arXiv:0710.2783 [astro-ph].

\bibitem{Einasto}
M.~Einasto, J.~Einasto, E.~Tago, G.~B.~Dalton and H.~Andernach,
``The Structure of the Universe Traced by Rich Clusters of Galaxies'',
Mon.\ Not.\ Roy.\ Astron.\ Soc.\ {\bf 269} (1994) 301

\bibitem{Rudnick:2007kw}
  L.~Rudnick, S.~Brown and L.~R.~Williams,
  ``Extragalactic Radio Sources and the WMAP Cold Spot'',
  Astrophys.\ J.\  {\bf 671} (2007) 40
  [arXiv:0704.0908 [astro-ph]].

\bibitem{Copeland:2006wr}
  E.~J.~Copeland, M.~Sami and S.~Tsujikawa,
  ``Dynamics of dark energy'',
  Int.\ J.\ Mod.\ Phys.\ D {\bf 15} (2006) 1753
  [arXiv:hep-th/0603057].

\bibitem{Straumann:2006tv}
  N.~Straumann,
   ``Dark energy: Recent developments'',
  Mod.\ Phys.\ Lett.\ A {\bf 21} (2006) 1083
  [arXiv:hep-ph/0604231].

\bibitem{Sahni:2006pa}
  V.~Sahni and A.~Starobinsky,
  ``Reconstructing dark energy'',
  Int.\ J.\ Mod.\ Phys.\  D {\bf 15} (2006) 2105
  [arXiv:astro-ph/0610026].

\bibitem{Riess:2004nr}
  A.~G.~Riess {\it et al.}  [Supernova Search Team Collaboration],
  ``Type Ia Supernova Discoveries at $z>1$ From the Hubble Space Telescope: Evidence for Past Deceleration and Constraints on Dark Energy Evolution'',
  Astrophys.\ J.\  {\bf 607} (2004) 665
  [arXiv:astro-ph/0402512].

\bibitem{Riess:2006fw}
  A.~G.~Riess {\it et al.},
  ``New Hubble Space Telescope Discoveries of Type Ia Supernovae at $z > 1$: Narrowing Constraints on the Early Behavior of Dark Energy'',
  Astrophys.\ J.\  {\bf 659} (2007) 98
  [arXiv:astro-ph/0611572].

\bibitem{Eisenstein:2005su}
  D.~J.~Eisenstein {\it et al.},
   ``Detection of the Baryon Acoustic Peak in the Large-Scale Correlation Function of SDSS Luminous Red Galaxies'',
  Astrophys.\ J.\  {\bf 633} (2005) 560
  [arXiv:astro-ph/0501171].

\bibitem{Spergel:2006hy}
  D.~N.~Spergel {\it et al.}  [WMAP Collaboration],
  ``Wilkinson Microwave Anisotropy Probe (WMAP) three year results: Implications for cosmology'',
  Astrophys.\ J.\ Suppl.\  {\bf 170} (2007) 377
  [arXiv:astro-ph/0603449].

\bibitem{Schwarz:2002ba}
  D.~J.~Schwarz,
  ``Accelerated expansion without dark energy'',
  arXiv:astro-ph/0209584.

\bibitem{Wetterich:2001kr}
  C.~Wetterich,
  ``Can structure formation influence the cosmological evolution?'',
  Phys.\ Rev.\  D {\bf 67} (2003) 043513
  [arXiv:astro-ph/0111166].

\bibitem{Rasanen:2003fy}
  S.~R\"as\"anen,
  ``Dark energy from backreaction'',
  JCAP {\bf 0402} (2004) 003
  [arXiv:astro-ph/0311257].

\bibitem{Rasanen:2006kp}
  S.~R\"as\"anen,
  ``Accelerated expansion from structure formation'',
  JCAP {\bf 0611} (2006) 003
  [arXiv:astro-ph/0607626].

\bibitem{Rasanen:2008it}
  S.~R\"as\"anen,
  ``Evaluating backreaction with the peak model of structure formation'',
  JCAP {\bf 0804} (2008) 026
  [arXiv:0801.2692 [astro-ph]].

\bibitem{Rasanen:2008be}
  S.~R\"as\"anen,
  ``Light propagation in statistically homogeneous and isotropic dust universes'',
  JCAP {\bf 0902} (2009) 011
  [arXiv:0812.2872 [astro-ph]].

\bibitem{Buchert:2007ik}
  T.~Buchert,
  ``Dark Energy from Structure - A Status Report'',
  Gen.\ Rel.\ Grav.\  {\bf 40} (2008) 467
  [arXiv:0707.2153 [gr-qc]].

\bibitem{Ishibashi:2005sj}
  A.~Ishibashi and R.~M.~Wald,
  ``Can the acceleration of our universe be explained by the effects of inhomogeneities?'',
  Class.\ Quant.\ Grav.\  {\bf 23} (2006) 235
  [arXiv:gr-qc/0509108].

\bibitem{Paranjape:2008jc}
  A.~Paranjape and T.~P.~Singh,
  ``Cosmic Inhomogeneities and the Average Cosmological Dynamics'',
  Phys.\ Rev.\ Lett.\  {\bf 101} (2008) 181101
  [arXiv:0806.3497 [astro-ph]].

\bibitem{Kolb:2008bn}
  E.~W.~Kolb, V.~Marra and S.~Matarrese,
  ``On the description of our cosmological spacetime as a perturbed conformal Newtonian metric and implications for the backreaction proposal for the accelerating universe'',
  Phys.\ Rev.\  D {\bf 78} (2008) 103002
  [arXiv:0807.0401 [astro-ph]].

\bibitem{Shirokov}
 M.~F.~Shirokov and I.~Z.~Fisher,
 ``Isotropic Space with Discrete Gravitational-Field Sources. On the Theory of a Nonhomogeneous Isotropic Universe'',
 Soviet Ast.\ (1963) 6, 699

\bibitem{Ell84}
G.~F.~R. Ellis,
``Relativistic cosmology: its nature, aims and problems'', p. 215 in {\em General Relativity and Gravitation},
edited by B.~Bertotti, F.~de~Felice, \& A.~Pascolini, D. Reidel Publishing Company, 1984.

\bibitem{EllisStoeger}
 G.~F.~R.~Ellis and W.~Stoeger
 ``The 'fitting problem' in cosmology'',
 1987 Class. Quant. Grav. {\bf 4} 1697

\bibitem{Ellis:1999sx}
  G.~F.~R.~Ellis,
  ``83 years of general relativity and cosmology: Progress and problems'',
  Class.\ Quant.\ Grav.\  {\bf 16} (1999) A37.

\bibitem{Ellis:2005uz}
  G.~F.~R.~Ellis and T.~Buchert,
  ``The universe seen at different scales'',
  Phys.\ Lett.\ A {\bf 347} (2005) 38
  [arXiv:gr-qc/0506106].

\bibitem{Buchert:1999er}
  T.~Buchert,
   ``On average properties of inhomogeneous fluids in general relativity. I: Dust cosmologies'',
  Gen.\ Rel.\ Grav.\  {\bf 32} (2000) 105
  [arXiv:gr-qc/9906015].

\bibitem{Mattsson:2007qp}
  T.~Mattsson and M.~Ronkainen,
  ``Exploiting scale dependence in cosmological averaging'',
  JCAP {\bf 0802} (2008) 004
  [arXiv:0708.3673 [astro-ph]].

\bibitem{Apostolopoulos:2006eg}
  P.~S.~Apostolopoulos, N.~Brouzakis, N.~Tetradis and E.~Tzavara,
  ``Cosmological acceleration and gravitational collapse'',
  JCAP {\bf 0606} (2006) 009
  [arXiv:astro-ph/0603234].

\bibitem{Kai:2006ws}
  T.~Kai, H.~Kozaki, K.~I.~Nakao, Y.~Nambu and C.~M.~Yoo,
  ``Can inhomogeneties accelerate the cosmic volume expansion?'',
  Prog.\ Theor.\ Phys.\  {\bf 117} (2007) 229
  [arXiv:gr-qc/0605120].

\bibitem{Paranjape:2006cd}
  A.~Paranjape and T.~P.~Singh,
  ``The Possibility of Cosmic Acceleration via Spatial Averaging in Lema\^itre-Tolman-Bondi Models'',
  Class.\ Quant.\ Grav.\  {\bf 23} (2006) 6955
  [arXiv:astro-ph/0605195].

\bibitem{Wiltshire:2007jk}
  D.~L.~Wiltshire,
  ``Cosmic clocks, cosmic variance and cosmic averages'',
  New J.\ Phys.\  {\bf 9} (2007) 377
  [arXiv:gr-qc/0702082].

\bibitem{Wiltshire:2007fg}
  D.~L.~Wiltshire,
  ``Exact solution to the averaging problem in cosmology'',
  Phys.\ Rev.\ Lett.\  {\bf 99} (2007) 251101
  [arXiv:0709.0732 [gr-qc]].

\bibitem{Leith:2007ay}
  B.~M.~Leith, S.~C.~C.~Ng and D.~L.~Wiltshire,
  ``Gravitational energy as dark energy: Concordance of cosmological tests'',
  Astrophys.\ J.\  {\bf 672} (2008) L91
  [arXiv:0709.2535 [astro-ph]].

\bibitem{Wiltshire:2007zh}
  D.~L.~Wiltshire,
  ``Gravitational energy and cosmic acceleration'',
  Int.\ J.\ Mod.\ Phys.\  D {\bf 17} (2008) 641
  [arXiv:0712.3982 [gr-qc]].

\bibitem{Wiltshire:2007zj}
  D.~L.~Wiltshire,
  ``Dark energy without dark energy'',
  arXiv:0712.3984 [astro-ph].

\bibitem{Wiltshire:2008sg}
  D.~L.~Wiltshire,
  ``Cosmological equivalence principle and the weak-field limit'',
  Phys.\ Rev.\  D {\bf 78} (2008) 084032
  [arXiv:0809.1183 [gr-qc]].

\bibitem{Smith:2008tc}
  K.~M.~Smith and D.~Huterer,
  ``No evidence for the cold spot in the NVSS radio survey'',
  arXiv:0805.2751 [astro-ph].

\bibitem{Moffat:1994qy}
  J.~W.~Moffat and D.~C.~Tatarski,
  ``Cosmological observations in a local void'',
  arXiv:astro-ph/9407036.

\bibitem{Zehavi:1998gz}
  I.~Zehavi, A.~G.~Riess, R.~P.~Kirshner and A.~Dekel,
  ``A Local Hubble Bubble from SNe Ia?'',
  Astrophys.\ J.\  {\bf 503} (1998) 483
  [arXiv:astro-ph/9802252].

\bibitem{Tomita:2000jj}
  K.~Tomita,
  ``A Local Void and the Accelerating Universe'',
  Mon.\ Not.\ Roy.\ Astron.\ Soc.\  {\bf 326} (2001) 287
  [arXiv:astro-ph/0011484].

\bibitem{Frith:2005et}
  W.~J.~Frith, N.~Metcalfe and T.~Shanks,
  ``New H-band Galaxy Number Counts: A Large Local Hole in the Galaxy Distribution?'',
  Mon.\ Not.\ Roy.\ Astron.\ Soc.\  {\bf 371} (2006) 1601
  [arXiv:astro-ph/0509875].

\bibitem{Lemaitre:1933qe}
  G.~Lema\^itre,
  Annales Soc.\ Sci.\ Brux.\ Ser.\ I Sci.\ Math.\ Astron.\ Phys.\ A {\bf 53} (1933) 51.

  For an English translation, see:

  G.~Lema\^itre,
  ``The Expanding Universe'',
  Gen.\ Rel.\ Grav.\  {\bf 29} (1997) 641.

\bibitem{Plebanski:2006sd}
  J.~Plebanski and A.~Krasinski,
  ``An Introduction to General Relativity and Cosmology'',
Cambridge University Press (2006).

\bibitem{Mustapha:1998jb}
  N.~Mustapha, C.~Hellaby and G.~F.~R.~Ellis,
   ``Large scale inhomogeneity versus source evolution: Can we distinguish them observationally?'',
  Mon.\ Not.\ Roy.\ Astron.\ Soc.\  {\bf 292} (1997) 817
  [arXiv:gr-qc/9808079].

\bibitem{Celerier:1999hp}
  M.~N.~Celerier,
  ``Do we really see a cosmological constant in the supernovae data ?'',
  Astron.\ Astrophys.\  {\bf 353} (2000) 63
  [arXiv:astro-ph/9907206].

\bibitem{Alnes:2005rw}
  H.~Alnes, M.~Amarzguioui and {\O}.~Gr{\o}n,
  ``An inhomogeneous alternative to dark energy?'',
  Phys.\ Rev.\ D {\bf 73} (2006) 083519
  [arXiv:astro-ph/0512006].

\bibitem{Enqvist:2006cg}
  K.~Enqvist and T.~Mattsson,
  ``The effect of inhomogeneous expansion on the supernova observations'',
  JCAP {\bf 0702} (2007) 019
  [arXiv:astro-ph/0609120].

\bibitem{Iguchi:2001sq}
  H.~Iguchi, T.~Nakamura and K.~I.~Nakao,
  ``Is dark energy the only solution to the apparent acceleration of the present universe?'',
  Prog.\ Theor.\ Phys.\  {\bf 108} (2002) 809
  [arXiv:astro-ph/0112419].

\bibitem{Biswas:2006ub}
  T.~Biswas, R.~Mansouri and A.~Notari,
  ``Nonlinear Structure Formation and Apparent Acceleration: an Investigation'',
  JCAP {\bf 0712} (2007) 017
  [arXiv:astro-ph/0606703].

\bibitem{Tanimoto:2007dq}
  M.~Tanimoto and Y.~Nambu,
  ``Luminosity distance-redshift relation for the LTB solution near the center'',
  Class.\ Quant.\ Grav.\  {\bf 24} (2007) 3843
  [arXiv:gr-qc/0703012].

\bibitem{Alexander:2007xx}
  S.~Alexander, T.~Biswas, A.~Notari and D.~Vaid,
  ``Local Void vs Dark Energy: Confrontation with WMAP and Type Ia Supernovae'',
  arXiv:0712.0370 [astro-ph].

\bibitem{GarciaBellido:2008nz}
  J.~Garcia-Bellido and T.~Haugboelle,
  ``Confronting Lema\^itre-Tolman-Bondi models with Observational Cosmology'',
  JCAP {\bf 0804} (2008) 003
  [arXiv:0802.1523 [astro-ph]].

\bibitem{GarciaBellido:2008yq}
  J.~Garcia-Bellido and T.~Haugboelle,
  ``The radial BAO scale and Cosmic Shear, a new observable for Inhomogeneous Cosmologies'',
  arXiv:0810.4939 [astro-ph].

\bibitem{Zibin:2008vj}
  J.~P.~Zibin,
  ``Scalar Perturbations on Lema\^itre-Tolman-Bondi Spacetimes'',
  Phys.\ Rev.\  D {\bf 78} (2008) 043504
  [arXiv:0804.1787 [astro-ph]].

\bibitem{Alnes:2006pf}
  H.~Alnes and M.~Amarzguioui,
  ``CMB anisotropies seen by an off-center observer in a spherically symmetric inhomogeneous universe'',
  Phys.\ Rev.\  D {\bf 74} (2006) 103520
  [arXiv:astro-ph/0607334].

\bibitem{Pain:2002wj}
  R.~Pain {\it et al.}  [Supernova Cosmology Project Collaboration],
  ``The distant Type Ia supernova rate'',
  Astrophys.\ J.\  {\bf 577} (2002) 120
  [arXiv:astro-ph/0205476].

\bibitem{Tonry:2004}
J.~L.~Tonry
``Supernovae and Dark Energy'',
Physica Scripta. Vol. T117, 11-16, 2005.

\bibitem{Weinberg1}
 S.~Weinberg
 ``Apparent luminosities in a locally inhomogeneous universe'',
 Astrophys.\ J.\ {\bf 208} (1976), p. L1-L3.

\bibitem{Ellis:1998ha}
  G.~F.~R.~Ellis, B.~A.~Bassett and P.~K.~S.~Dunsby,
  ``Lensing and caustic effects on cosmological distances'',
  Class.\ Quant.\ Grav.\  {\bf 15} (1998) 2345
  [arXiv:gr-qc/9801092].

\bibitem{Linder1}
  E.~V.~Linder,
  ``Light propagation in generalized Friedmann universes'',
  Astron.\ Astrophys.\ {\bf 206} (1988) 190

\bibitem{Kantowski:1995bd}
  R.~Kantowski, T.~Vaughan and D.~Branch,
  ``The Effects of Inhomogeneities on Evaluating the Deceleration Parameter $q_0$'',
  Astrophys.\ J.\  {\bf 447} (1995) 35
  [arXiv:astro-ph/9511108].

\bibitem{Kantowski:1998ju}
  R.~Kantowski,
  ``The Effects of Inhomogeneities on Evaluating the mass parameter $\Omega_m$ and the cosmological constant $\Lambda$'',
  arXiv:astro-ph/9802208.

\bibitem{Kantowski:2003yt}
  R.~Kantowski,
  ``The Lame$^{\prime}$ Equation for Distance-Redshift in Partially Filled Beam Friedmann-Lema\^\i tre-Robertson-Walker Cosmology'',
  Phys.\ Rev.\  D {\bf 68} (2003) 123516
  [arXiv:astro-ph/0308419].

\bibitem{Sachs}
  R.~K.~Sachs,
  ``Gravitational Waves in General Relativity. VI. The Outgoing Radiation Condition'',
  Proc.\ R.\ Soc.\ London A {\bf 264} (1961) 309.

\bibitem{Etherington}
  I.~M.~H.~Etherington,
  ``On the definition of distance in general relativity'',
  Phil.\ Mag.\ ser.\ 7, {\bf 15} (1933) 761.

\bibitem{Ellis}
  G.~F.~R.~Ellis, ``Relativistic Cosmology'', p. 104 in Proc. School Enrico Fermi, ``General Relativity and Cosmology'', Ed.~R.~K.~Sachs, Academic Press (New York 1971).

\bibitem{Santos:2008tz}
  R.~C.~Santos and J.~A.~S.~Lima,
  ``Clustering, Angular Size and Dark Energy'',
  Phys.\ Rev.\  D {\bf 77} (2008) 083505
  [arXiv:0803.1865 [astro-ph]].

\bibitem{Peacock}
J.~A.~Peacock, ``Cosmological Physics'', Cambridge University Press (1999).

\bibitem{Springel:2005nw}
  V.~Springel {\it et al.},
  ``Simulating the joint evolution of quasars, galaxies and their large-scale distribution'',
  Nature {\bf 435} (2005) 629
  [arXiv:astro-ph/0504097].

\bibitem{Sarkar:2007cx}
  S.~Sarkar,
  ``Is the evidence for dark energy secure?'',
  Gen.\ Rel.\ Grav.\  {\bf 40} (2008) 269
  [arXiv:0710.5307 [astro-ph]].

\bibitem{Blanchard:2003du}
  A.~Blanchard, M.~Douspis, M.~Rowan-Robinson and S.~Sarkar,
  ``An alternative to the cosmological 'concordance model' '',
  Astron.\ Astrophys.\  {\bf 412} (2003) 35
  [arXiv:astro-ph/0304237].

\bibitem{Blanchard:2005ev}
  A.~Blanchard, M.~Douspis, M.~Rowan-Robinson and S.~Sarkar,
  ``Large-scale galaxy correlations as a test for dark energy'',
  Astron.\ Astrophys.\  {\bf 449} (2006) 925
  [arXiv:astro-ph/0512085].

\bibitem{Hunt:2004vt}
  P.~Hunt and S.~Sarkar,
  ``Multiple inflation and the WMAP 'glitches' '',
  Phys.\ Rev.\  D {\bf 70} (2004) 103518
  [arXiv:astro-ph/0408138].

\bibitem{Hunt:2007dn}
  P.~Hunt and S.~Sarkar,
  ``Multiple inflation and the WMAP 'glitches' II. Data analysis and cosmological parameter extraction'',
  Phys.\ Rev.\  D {\bf 76} (2007) 123504
  [arXiv:0706.2443 [astro-ph]].

\bibitem{Hunt:2008wp}
  P.~Hunt and S.~Sarkar,
  ``Constraints on large scale voids from WMAP-5 and SDSS'',
  arXiv:0807.4508 [astro-ph].

\bibitem{Adams:1997de}
  J.~A.~Adams, G.~G.~Ross and S.~Sarkar,
  ``Multiple inflation'',
  Nucl.\ Phys.\  B {\bf 503} (1997) 405
  [arXiv:hep-ph/9704286].

\bibitem{Yao}
  W.~M.~Yao {\it et al.},
  `` Review of Particle Physics'',
  [Particle Data Group], J.~Phys.~G {\bf 33}, 1 (2006).

\bibitem{Drexlin:2005zt}
  G.~Drexlin  [KATRIN Collaboration],
  ``KATRIN: Direct measurement of a sub-eV neutrino mass'',
  Nucl.\ Phys.\ Proc.\ Suppl.\  {\bf 145} (2005) 263.

\bibitem{Elgaroy:2003yh}
  {\O}.~Elgar{\o}y and O.~Lahav,
  ``The role of priors in deriving upper limits on neutrino masses from the 2dFGRS and WMAP'',
  JCAP {\bf 0304} (2003) 004
  [arXiv:astro-ph/0303089].

\bibitem{Singh:2002de}
  S.~Singh and C.~P.~Ma,
  ``Neutrino clustering in cold dark matter halos: Implications for ultra high energy cosmic rays'',
  Phys.\ Rev.\  D {\bf 67} (2003) 023506
  [arXiv:astro-ph/0208419].

\bibitem{Hu:1995en}
  W.~Hu and N.~Sugiyama,
  ``Small scale cosmological perturbations: An Analytic approach'',
  Astrophys.\ J.\  {\bf 471} (1996) 542
  [arXiv:astro-ph/9510117].

\bibitem{Rakic:2007ve}
  A.~Rakic and D.~J.~Schwarz,
  ``Correlating anomalies of the microwave sky: The Good, the Evil and the Axis'',
  Phys.\ Rev.\  D {\bf 75} (2007) 103002
  [arXiv:astro-ph/0703266].

\bibitem{Gurzadyan:2007ic}
  V.~G.~Gurzadyan, A.~Kashin, C.~L.~Bianco, H.~Khachatryan and G.~Yegorian,
  ``On Axial and Plane--Mirror Inhomogeneities in the WMAP3 Cosmic Microwave Background Maps'',
  Mod.\ Phys.\ Lett.\  A {\bf 22} (2007) 2955
  [arXiv:0709.0886 [astro-ph]].

\bibitem{Freedman:2000cf}
  W.~L.~Freedman {\it et al.},
  ``Final Results from the Hubble Space Telescope Key Project to Measure the Hubble Constant'',
  Astrophys.\ J.\  {\bf 553} (2001) 47
  [arXiv:astro-ph/0012376].

\bibitem{Sandage:2006cv}
  A.~Sandage, G.~A.~Tammann, A.~Saha, B.~Reindl, F.~D.~Macchetto and N.~Panagia,
  ``The Hubble Constant: A Summary of the HST Program for the Luminosity Calibration of Type Ia Supernovae by Means of Cepheids'',
  Astrophys.\ J.\  {\bf 653} (2006) 843
  [arXiv:astro-ph/0603647].

\bibitem{Hasinger:2002wg}
  G.~Hasinger, N.~Schartel and S.~Komossa,
  ``Discovery of an ionized Fe-K edge in the $z=3.91$ Broad Absorption Line Quasar APM $08279+5255$ with XMM-Newton'',
  Astrophys.\ J.\  {\bf 573} (2002) L77
  [arXiv:astro-ph/0207005].

\bibitem{Komossa:2002cn}
  S.~Komossa and G.~Hasinger,
  ``The X-ray evolving universe: (ionized) absorption and dust, from nearby Seyfert galaxies to high-redshift quasars'',
  arXiv:astro-ph/0207321.

\bibitem{Jain:2005gu}
  D.~Jain and A.~Dev,
  ``Age of High Redshift Objects - a Litmus Test for the Dark Energy Models'',
  Phys.\ Lett.\  B {\bf 633} (2006) 436
  [arXiv:astro-ph/0509212].

\bibitem{Fields:2006ga}
  B.~Fields and S.~Sarkar,
  ``Big-bang nucleosynthesis (PDG mini-review)'',
  arXiv:astro-ph/0601514.

\bibitem{Krauss}
L.~M.~Krauss and B.~Chaboyer,
``Age Estimates of Globular Clusters in the Milky Way: Constraints on Cosmology'',
Science {\bf 299} (2003) 65.

\bibitem{Wang:2006cq}
  X.~F.~Wang, L.~F.~Wang, R.~Pain, X.~Zhou and Z.~W.~Li,
  ``Determination of the Hubble constant, the intrinsic scatter of luminosities of Type Ia SNe, and evidence for non-standard dust in other galaxies'',
  Astrophys.\ J.\  {\bf 645} (2006) 488
  [arXiv:astro-ph/0603392].

\bibitem{Schwarz:2007wf}
  D.~J.~Schwarz and B.~Weinhorst,
  ``(An)isotropy of the Hubble diagram: comparing hemispheres'',
  arXiv:0706.0165 [astro-ph].

\bibitem{Seikel:2007pk}
  M.~Seikel and D.~J.~Schwarz,
  ``How strong is the evidence for accelerated expansion?'',
  JCAP {\bf 0802} (2008) 007
  [arXiv:0711.3180 [astro-ph]].

\bibitem{McClure:2007vv}
  M.~L.~McClure and C.~C.~Dyer,
  ``Anisotropy in the Hubble constant as observed in the HST Extragalactic Distance Scale Key Project results'',
  New Astron.\  {\bf 12} (2007) 533
  [arXiv:astro-ph/0703556].

\bibitem{Gurzadyan:2008va}
  V.~G.~Gurzadyan {\it et al.},
  ``Kolmogorov CMB Sky'',
  arXiv:0811.2732 [astro-ph].

\bibitem{Kutschera:2006bh}
  M.~Kutschera and M.~Dyrda,
  ``Coincidence of universe age in LambdaCDM and Milne cosmologies'',
  Acta Phys.\ Polon.\  B {\bf 38} (2007) 215
  [arXiv:astro-ph/0605175].

\end{thebibliography}
\end{document}